\documentclass{ws-rv9x6}
\usepackage{subfigure}   
\usepackage{ws-rv-thm}   
\usepackage{ws-rv-van}   
\makeindex

\newcommand{\eq}[1]{(\ref{#1})}
\newcommand{\fig}[1]{Fig.~\ref{#1}}

\newcommand{\be}{\begin{equation}}
\newcommand{\ee}{\end{equation}}

\newcommand\disp{\displaystyle}

\newcommand{\la}{\left<}
\newcommand{\ra}{\right>}
\newcommand{\eps}{\varepsilon}

\newcommand{\re}{\textrm{Re}\,}
\newcommand{\im}{\textrm{Im}\,}

\begin{document}

\chapter[Non-Euclidean geometry in nature]
{Non-Euclidean geometry in nature}\label{ra_ch1}

\author[S. Nechaev]{Sergei Nechaev$^{1,2}$}

\address{$^1$Interdisciplinary Scientific Center J.-V. Poncelet\footnote{International
Joint Research Unit (UMI 2615) CNRS -- IUM -- IITP RAS -- MI RAS -- Skoltech -- HSE, Moscow,
Russian Federation}, Moscow, Russia \\ $^2$P.N. Lebedev Physical Institute RAS, Moscow, Russia, \\
nechaev@lptms.u-psud.fr}

\begin{abstract}
I describe the manifestation of the non-Euclidean geometry in the behavior of collective observables of some complex physical systems. Specifically, I consider the formation of equilibrium shapes of plants and statistics of sparse random graphs. For these systems I discuss the following interlinked questions: (i) the optimal embedding of plants leaves in the three-dimensional space, (ii) the spectral statistics of sparse random matrix ensembles.
\end{abstract}
\body

\tableofcontents

\bigskip

\section{Non-Euclidean geometry in complex physical systems: Knots, jupe \`a godets,
ultrametricity and number theory}
\label{sec1}

Our everyday experience tells us that we live in a three-dimensional Euclidean space... well, in a
four-dimensional space-time, if we take into account the fourth dimension, the time. If
gravitational effects are supposed to be negligible, we expect that this four-dimensional
space-time, $X=\{x_1,x_2,x_3,x_4\}$, is flat and has Euclidean metric, $ds^2=\sum_{i=1}^4 dx_i^2$.
However, when looking at the world around us through the magnifying glass, one can notice various incarnations of the non-Euclidean Hyperbolic geometry, also known as the Lobachevsky geometry, in the behavior of collective observables of some complex systems. The Lobachevsky geometry emerges in bizarre buckling of some leafs of plants, in structure of cones and corals, in fashion of clothes, in folding of DNA in chromosomes, in statistics of eigenvalues of rare random clusters...

Despite all these systems are of very different physical origins, they still have some common property which is crucial for the nontrivial statistical behavior of collective variables: the systems considered above have exponential number of states, and these states should be embedded into the low-dimensional ``physical'' space subject to some natural constraints. What is seen as a result -- is an ``optimal'' (typical) statistical configuration which minimizes the ``conflicts'' imposed by constraints. Namely,
\begin{itemize}
\item[-] for a leaf of a plant, the exponential proliferation of cells is in conflict with the embedding of a growing tissue into the tree-dimensional space, thus leading to a peculiar stable fractal boundary structure of the plant;
\item[-] for packing of a DNA in chromosome, the selection of an unknotted compact loop among  exponentially large number of various topological states, leads to a stable space-filling hierarchical fractal folding of the entire molecule;
\item[-] for an ensemble of sparse adjacency matrices, the condition to have small fraction of non-zero elements in each row and column leads to a stable ultrametric structure of the eigenvalue density which is ultimately related to the energetic approach of compressed lattices considered in the context of phyllotaxis.
\end{itemize}

The review is structured as follows. The Section \ref{sec1} consists of four introductory parts \ref{sec11}-\ref{sec14}, where the problems and questions dealing with manifestation of Hyperbolic geometry in nature are formulated, while the detailed consideration of subjects mentioned in Subsections \ref{sec12}-\ref{sec14} are provided in Sections \ref{sec2} and \ref{sec3}.

\subsection{Topology of knots}
\label{sec11}

Unknotting heavily tangled ropes, we are not even surprised that long threads, left to themselves, get entangled in most the unpleasant way of all. We can only utter a bitter sigh, when almost unravelled ball of wool accidentally falls into the clutches of a curious kitten... and definitely, we are hardly aware that the spontaneous knotting of long strands, managed by the laws of probability theory, is a consequence of non-Euclidean geometry of the phase space of knots, i.e., a hypothetical space that contains all possible knots and in which there is a concept of metric, or the distance between "topologically similar" knots. The more the topological types of two knots differ, the greater the distance between them in this space. In general, the construction of this space, called the universal covering, and the description of its properties, are challenging problems. However, for some special cases, this space can be described analytically. So, it turns out that, in order to untie ropes in a purely scientific way, one has first to understand the Lobachevsky-Riemann geometry and its relation to knot theory, and then to learn how probability theory works in this non-Euclidean space.

Some conceptual steps in the analytic description of topological interactions, which constitute the basis of a new interdisciplinary branch in mathematical physics, emerging at the edge of topology and statistical physics of fluctuating non-phantom rope-like objects have been reviewed in Ref. [\refcite{india}]. This new branch is called statistical (or probabilistic) topology. Its most fascinating manifestation, in my opinion, is connected with the nonperturbative description of DNA packing in chromosomes in a form of a crumpled globule [\refcite{gns,gr-rab}]. After the experimental work [\refcite{mirny}] of the MIT-Harvard team in 2009, the concept of crumpled globule became a kind of a new paradigm allowing to understand the mathematical origin of many puzzling features of DNA structuring and functioning in the human genome. The mathematical background of the crumpled globule deals with the statistics of Brownian bridges in the non-Euclidean space of constant negative curvature (see Ref. [\refcite{india}] for review).

\begin{figure}[ht]
\centerline{\includegraphics[width=3.3cm]{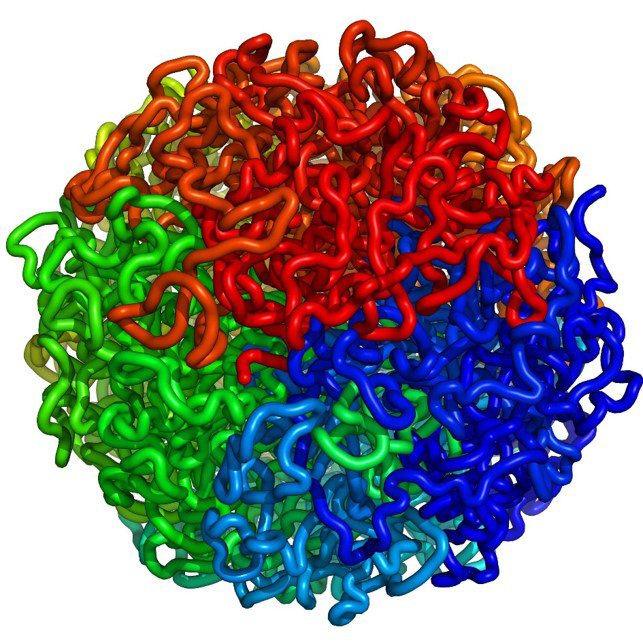} \hspace{3mm}
\includegraphics[width=3.3cm]{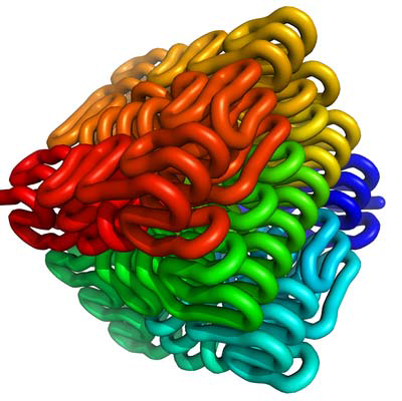} \includegraphics[width=4.3cm]{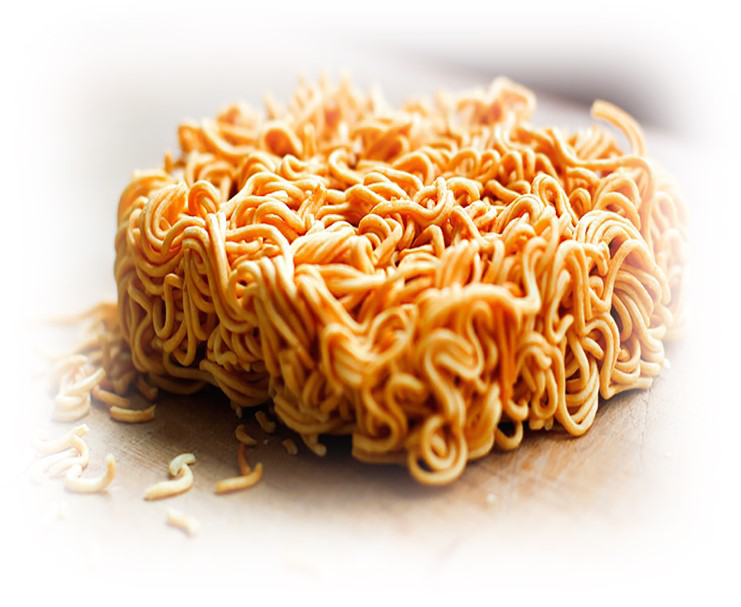}}
\caption{Crumpled polymer globule (left), three-dimensional Peano curve (center), brick
of dehydrated spaghetti (right).}
\label{fig01}
\end{figure}

The crumpled globule shown in \fig{fig01}a is a state of a polymer chain which in a wide range of
scales is self-similar and almost unknotted, forming a fractal space-filling structure. The chain
is painted from one end to another in continuously changing colors from red to violet. Such
painting allows to visualize the peculiar spatial structure of the globule: parts of the chain
which are neighbors along the path, are also neighbors in the space. This structure resembles the three-dimensional Peano curve depicted in \fig{fig01}b.

The self-similarity and the absence of knots, are essential for genome folding: fractal
organization makes genome tightly packed in a broad range of scales, while the lack of knots
ensures easy and independent opening and closing genomic domains, necessary for transcription
[\refcite{gr-rab,mirny2}]. In a three-dimensional space such a tight packing results in a
\emph{space-filling} with the fractal dimension $D_f=D=3$. The contact probability, $P_{i,j}$,
between two genomic units, $i$ and $j$ in a $N$-unit chain, depends on a combination of structural and energetic factors. Simple mean-field arguments (see, for example, Ref. [\refcite{mirny}]) demonstrate that in a fractal globule with $D_f=3$ the \emph{average} contact probability, $P(s)$, between two units separated by the genomic distance $s=|i-j|$, decays as
$$
{\cal P}(s) = (N-s)^{-1} \sum_{i=0}^{N-s} P_{i,i+s} \sim s^{-1}
$$
Combining the general hierarchical fractal globule folding mechanism with the assumption that
chromatin can be considered as a heteropolymer chain with a quenched primary sequence
[\refcite{filion}], one can reproduce the large-scale chromosome compartmentalization, not assumed explicitly from the very beginning. The account for the hierarchical folding together with the conjecture about quenched heterogeneous volume interactions [\refcite{hic-avet}] results in construction of artificial maps, possessing ultrametric properties (see the next Section for discussion of ultrametricity). The structures found in the frameworks of the toy model resemble the so-called Hi-C\footnote{Genome-whide chromosome conformational capture method} contact maps experimentally observed in chromosomes.

It should be mentioned that the crumpled globule resembles the Korean fast food "dosirak" shown in \fig{fig01}c which is the brick of dehydrated spaghetti. This resemblance has a deep physical sense: to absorb water effectively, spaghetti should unfold quickly as hot water penetrates inside the brick, to allow exterior folds open sequentially as fast as possible.

In this review I will not discuss anymore questions dealing with non-Euclidean geometry in statistics of knots, since this subject has deserved much attention in Ref. [\refcite{india}]. Rather I will concentrate on other challenging manifestations of hyperbolic geometry in physics where the "conflict" between the hyperbolic protocol of growth of natural tissues (like plants) with the restricted geometry of the three-dimensional space leads to high variety of self-similar shapes.

\subsection{Buckling of leaves}
\label{sec12}

\begin{figure}[ht]
\centerline{\includegraphics[width=5cm]{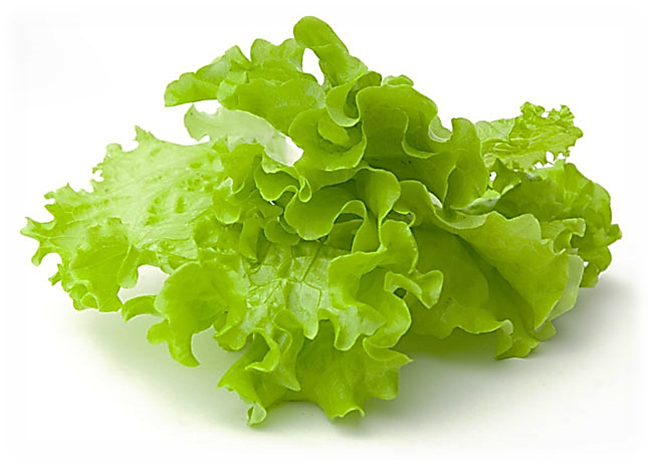} \includegraphics[width=4cm]{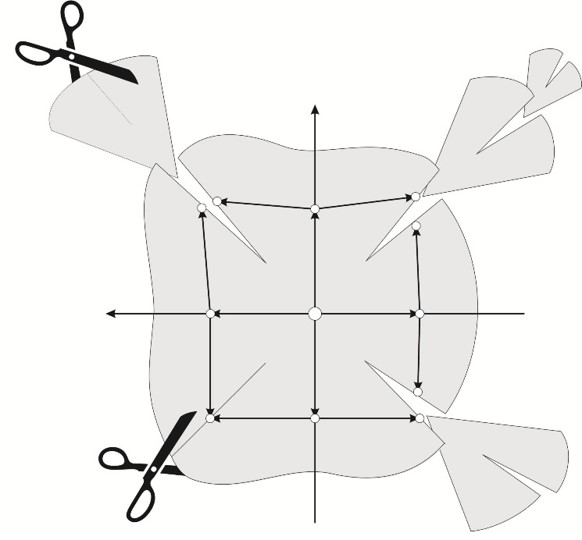} \hspace{1cm} \includegraphics[width=1.3cm]{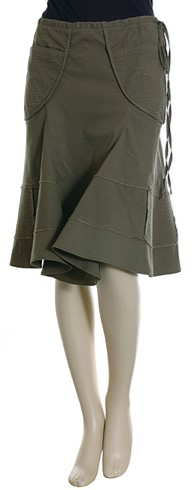}}
\caption{Leaf of a lettuce (left), construction of hyperbolic surface by "cutting and insertion"
procedure (center), jupe \`a godets (right).}
\label{fig02}
\end{figure}

Walking in the garden on in the forest we are often fascinated by diversity of shapes of buckled
and wrinkled leaves and flowers of plants. Such amusing variety of their forms has very
straightforward geometrical explanation. Buckling of the two-dimensional growing tissues emerges
due to the incompatibility of local internal (differential) growth protocol with geometric
constraints imposed by embedding of these tissues into the space. For example, buckling of a
lettuce leaf (see \fig{fig02}a) can be naively explained as a conflict between natural growth due
to the periphery cells division (typically, exponential), and growth of circumference of a planar
disc with gradually increasing radius. Due to a specific biological mechanism which inhibits growth of the cell experiencing sufficient external pressure, the division of inner cells is
insignificant, while periphery cells have less steric restrictions and proliferate easier. Thus,
the division of border cells has the major impact on the instabilities in a tissue. Such a
differential growth induces an increasing strain in a tissue near its edge and results in two
complimentary possibilities: i) in-plane tissue compression and/or redistribution of layer cells
accompanied by the in-plane circumference instability, or ii) out-of-plane tissue buckling with the formation of saddle-like surface regions. The latter is typical for various undulant negatively
curved shapes which are ubiquitous to many mild plants growing up in air or water where the gravity is of sufficiently small matter [\refcite{kelp,swinney}].

The circular surface whose perimeter grows exponentially with the radius is called hyperbolic.  Currently, two groups of works, devoted to the determination of shapes of growing tissues, can be distinguished in the literature. In the first group of works, denoted as ``geometric'' in Refs. [\refcite{voit,vas,borelli,polovnikov}], the determination of the typical profile of 2D surface is dictated by optimal "isometric" embedding of discretized 2D surface into 3D space, and is described by an appropriate change of conformal metrics.

In the second, more numerous group of works [\refcite{hebrew,lewicka,gemmer, marder,ignobel, swinney}], conventionally named ``energetic'', more realistic approach, based on a minimization of a bending free energy of non-uniformly deformed thin elastic plates, is realized. Authors of almost all works, devoted to buckling of 2D growing tissues, state that geometric and energetic approaches are complimentary to each other. Besides, still an explicit comparison of results of geometric and energetic approaches is very demanded.

Here we stay in the frameworks of the geometric approach. Take some elementary bounded domain of a flat surface, make finite radial cuts and insert in these cuts flat triangles, modelling the newly born periphery cells. The resulting surface obtained by joining together all elementary domains and inserted extra triangles, is not flat anymore. Continue the process recursively as shown in \fig{fig02}b, i.e. make cuts in the new surface, insert extra flat triangles and so on... One  gets this way a surface whose perimeter and area grow exponentially with the radius. Anyone who pays attention to the tendencies in fashion can immediately recognize in the growth protocol described above, the so-called "jupe \`a godets" shown in the \fig{fig02}c.

A similar problem had been discussed by the great the Russian mathematician, P.L. Chebyshev in his talk "On the cut of clothes" in Paris in August of 1878. Some traces concerning this talk could be found in Ref. [\refcite{cheb}], however apparently the entire work has remained unpublished. Among the modern developments of that subject one can mention the paper [\refcite{bowers}] by P. Bowers proving in 1997 the theorem on quasi-isometric embedding of a uniform binary tree into a negatively curved space, as well as the contribution [\refcite{duval}] by S. Duval and M. Tajine devoted to isometric embeddings of trees into metric spaces for fractal description. Despite rather transparent geometrical image, the very problem under consideration is still too vague. Let us formulate it in more rigorous terms, which allow for the mathematical analysis.

\begin{figure}[ht]
\centerline{\includegraphics[width=8cm]{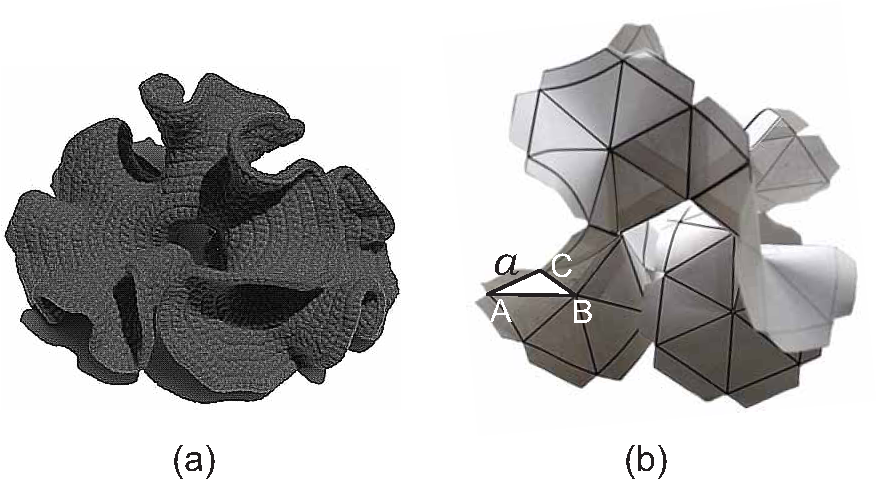}}
\caption{(a) Hyperbolic surface obtained by spiral crocheting from the center; (b) Hyperbolic
piecewise surface constructed by joining 7 equilateral flat triangles (copies of the triangle
$ABC$) in each vertex. The triangle $ABC$ is lying in $z = x+iy$ plane in the 3D Euclidean space,
$|AB|=|AC|=|BC|=a$.}
\label{fig03}
\end{figure}

To make our viewpoint more transparent, suppose that all cells in a growing colony, represented by equilateral triangles, divide independently and their proliferation is initiated by the first "protocell". Connecting the centers of neighboring triangles by nodes, we rise a graph $\gamma$. The number of vertices, $P_{\gamma}(k)$, in the generation $k$, grows exponentially with $k$: $P_{\gamma}(k) \sim c^k$ ($c>1$). It is known that the exponential graphs possess hyperbolic metrics, meaning that they can be isometrically (with fixed branch lengths and angles between adjacent branches) embedded into a hyperbolic plane. Thus, it is clear, that the corresponding surface, pulled on the isometry of such graph in the 3D Euclidean space, should be negatively curved. To have a relevant image [\refcite{crochet}], suppose that we grow the surface by crocheting it spirally starting from the center. Demanding two nearest neighboring circumference layers, $P(r)$ and $P(r+\Delta r)$, to differ by a factor of 2, i.e., $P(r+\Delta r)/P(r)=2$, we construct an exponentially growing surface -- see the \fig{fig03}a known in geometry as Amsler surface [\refcite{amsler}].

The crocheting surface has well-posed properties on large scales, but should be precisely described on the scale of order of the elementary cell. As we have mentioned, the microscopic description is
connected with the specific local growth protocol. The simplest way to generate the discrete
hyperbolic-like surface out of equilateral triangles, consists in gluing 7 such triangles in each
graph vertex and construct a piecewise surface $T$, shown in the \fig{fig03}b. On the scale less
than the elementary cell $ABC$ this surface is flat. Thus, the size $a$ ($|AB|=|AC|=|BC|=a$) of the triangle $ABC$ stands for the rigidity parameter, playing the role of a characteristic scale in our
problem, on which no deviations from the Euclidean metrics survive. Later on we shall see that
bending ability of growing surface crucially depends on this parameter: above some critical
$a_{cr}$ buckling is suppressed, while below $a_{cr}$ the tissue is flexible enough and buckling is profound.

\subsection{Ultrametricity}
\label{sec13}

Embedding of a Cayley tree ${\cal C}$ into the metric space is called "isometric" if ${\cal C}$
covers that space, preserving all angles and distances. The Cayley tree ${\cal C}$ isometrically
covers the surface of constant negative curvature (the Lobachevsky plane) ${\cal H}$. One of
possible representations of ${\cal H}$, known as a Poincar\'e model, is the upper half-plane $\im
z>0$ of the complex plane $z=x+iy$ endowed with the hyperbolic metric $ds^2=\frac{dx^2+dy^2}
{y^2}$. In Ref. [\refcite{vas}] we have constructed the "continuous" analog of the standard 3-branching Cayley tree by means of modular functions and have analyzed the structure of the barriers separating the neighboring valleys. In particular, we have shown that these barriers are
ultrametrically organized.

A metric space is a set of elements equipped by pairwise distances. Generically, the metric
$d(x,y)$ meets three requirements:
\begin{itemize}
\item[i)] Non-negativity, $d(x,y)>0$ for $x\neq y$, and $d(x,y)=0$ for $x=y$
\item[ii)] Symmetry, $d(x,y)=d(y,x)$
\item[iii)] the triangle inequality, $d(x,z)\le d(x,y) + d(y,z)$.
\end{itemize}
The concept of ultrametricity is related to a special class of metrics, obeying the strong triangular inequality, $d(x,z) \le \max\{d(x,y), d(y,z)\}$, which allows the existence of acute isosceles and equilateral triangles \emph{only}.

The strong triangle inequality radically changes the properties of spaces in comparison with our
intuitive expectations based of Euclidean geometry. In ultrametric spaces, the sum of distances
does not exceed the largest summand, i.e. the Archimedean  principle does not hold. Few
counterintuitive properties are related to ultrametric balls: every point inside a ball is its
center; the radius of a ball is equal to its diameter; there are no balls that overlap only
partially, if two balls have a common point, then one of them is completely embedded into the
other. The last property implies that any ultrametric ball can be divided into a set of smaller
balls, each of them can be divided into even smaller ones, etc. In other words, an ultrametric
space resembles not a line of segments, but a branching tree of hierarchically nested balls
(``basins"). The terminal points of a tree, constituting the tree boundary, also obeys the strong
triangle inequality: for any two points, $A$ and $B$, the distance to the root of the minimal subtree to which these points belong, determines the distance between $A$ and $B$. As a result, ultrametric geometry, in general, fixes taxonomic (i.e. hierarchical) tree-like relationships between the elements and, speaking figuratively, is closer to Lobachevsky geometry, rather to the Euclidean one.

Perhaps most naturally ultrametric spaces emerge in the realm of complex systems. Indeed, the
ultrametric description is inherently multi-scale, thus it is not surprising that ultrametricity
provides a fertile framework for describing complex systems characterized by a large number of
order parameters. For instance, signatures of ultrametricity were discovered in the context of spin glasses (see Ref. [\refcite{SG:ultra}] for an earlier review). Spin glasses are characterized by a huge number of metastable states and the equilibration in spin glasses proceeds via tree-like splitting of the phase space into hierarchically nested domains. When the temperature decreases, the most distant groups of spin-glass states get factorized first. Then, each of these groups splits into a number of subgroups, etc. The scales of hierarchically nested sets of equilibrated states satisfy the strong triangle inequality, i.e. the multi-scale splitting of spin-glass phases obey ultrametric relations.

Shortly after the formulation of the ultrametric ansatz in the context of spin glasses, similar
ideas were proposed for the description of native states of protein molecules [\refcite{nat_prot1, nat_prot2}]. Proteins are highly frustrated polymer systems characterized by rugged energy
landscapes of very high dimensionality. The ultrametricity of proteins implies a multi-scale
representation of the protein energy landscape by means of ranging the local minima into
hierarchically embedded basins of minima. Ultrametricity follows from the conjecture that the
equilibration time within the basin is significantly smaller than the time to escape the basin. As a result, the transitions between local minima obey the strong triangle inequality. In such an
approximation, dynamics on a complex energy landscape can be modelled via a jump-like random
process, propagating in an ultrametric space of states. The approximation of the protein energy
landscape by a self-similarly branching tree of basins has been shown to be very fruitful for
describing a large body of experimentally justified features of protein conformational dynamics and protein fluctuation mobility at temperatures covering very wide range from room temperatures up to
the deeply frozen states [\refcite{avet}].

Ultrametricity emerges in diverse subjects, e.g., in genetic trees, in the behavior of hierarchically organized ecological, social, and economic systems, etc. The fact that ultrametricity is naturally rooted in high dimensionality, randomness, and sparse statistics, has
been unambiguously shown recently in Ref. [\refcite{zubarev}]. It was proven that in a $D$-dimensional Euclidean space the distances between points in a highly sparse sampling tends to the ultrametric distances as $D\to\infty$.

Here we review another intriguing connection between rare-event statistics, ultrametricity and hyperbolic geometry. Consider a large $N\times N$ symmetric matrix $A$. The matrix elements, $a_{ij}=a_{ji}$, are assumed to be independent, identically distributed, binary random variables, such that diagonal elements vanish, $a_{ii}=0$, and
$$
a_{ij}=
\begin{cases}
1 & \mbox{with probability $q$} \\
0 & \mbox{with probability $1-q$}
\end{cases}
$$
for $i\ne j$. The matrix $A$ can be regarded as an adjacency matrix of a random Erd\H{o}s-R\'enyi
graph, ${\cal G}$, without self-connections. The eigenvalues, $\lambda_n$ ($n=1,...,N$), of the
symmetric  matrix $A$ are real. Let $\rho(\lambda)$ be the eigenvalue density of the ensemble of
such matrices. For $N\gg 1$ the limiting shape of $\rho(\lambda)$ is known in various cases. If
$q=\mathcal{O}(1)$, the function $\rho(\lambda)$ tends to the Wigner semicircle law, $\sqrt{4N-\lambda^2}$, typical for the Gaussian matrix ensembles. For $q=c/N$ ($c>1$), the matrix
$A$ is \emph{sparse} and the density $\rho(\lambda)$ of the sparse matrix ensemble has singularities at finite values of $c$ (see Refs. [\refcite{rod1,rod2,fyod}]). In Refs. [\refcite{evan,bauer,sem,kuch}], the behavior of the spectral density, $\rho(\lambda)$, has been analyzed near $c=1$ ($N\gg 1$). It has been pointed out that the function $\rho(\lambda)$ becomes more and more singular as $c$ approaches 1 from above.

\begin{figure}[ht]
\centerline{\includegraphics[width=4cm]{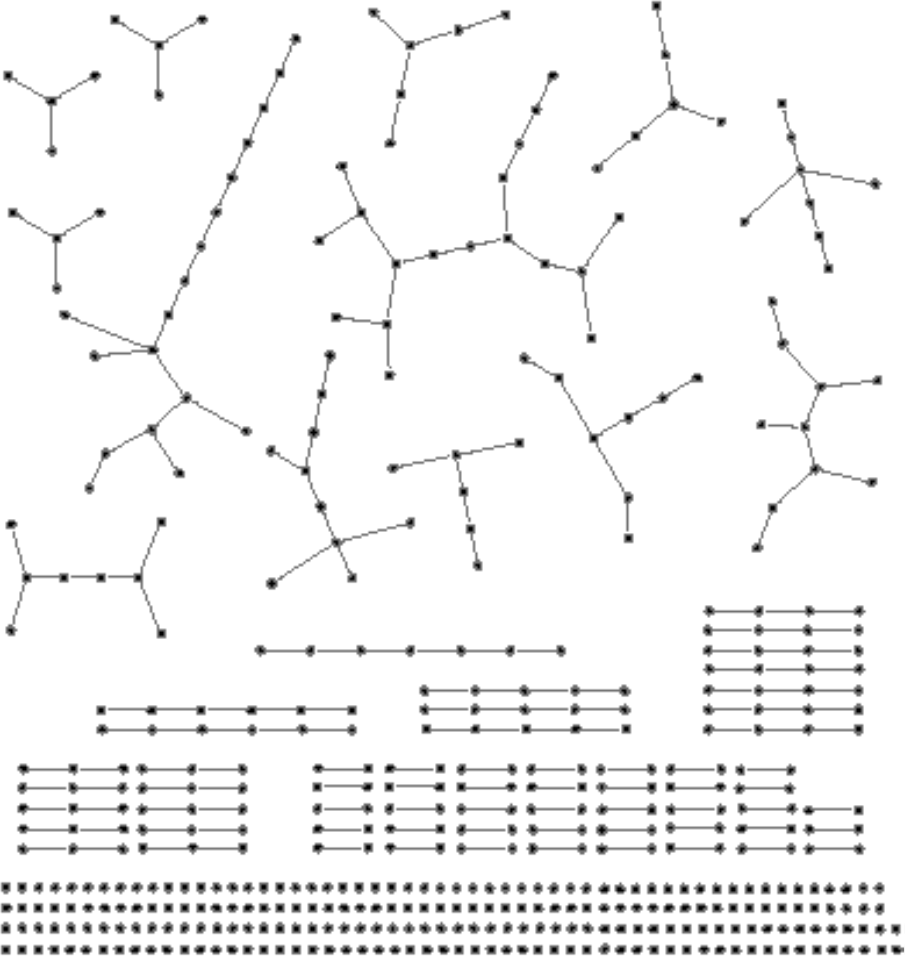} \hspace{5mm} \includegraphics[width=5.5cm]{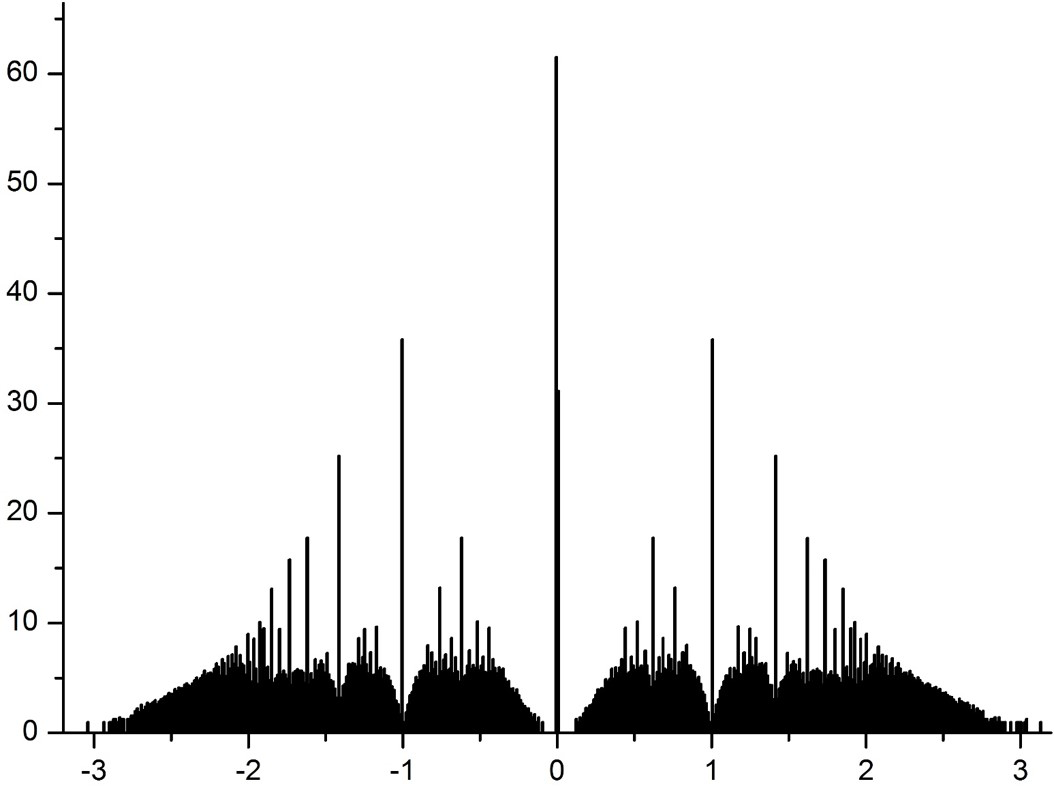}}
\caption{(left) Typical sample of components constituting a random graph associated with an adjacency matrix $A$, the parameters are $N=500$ and $q=1/N$; (right) the plot of $\rho(\lambda)$ shows the ultrametric structure of the spectral density and terminates close to the maximal eigenvalue for 3-branching trees, $|\lambda_{\max}^{\rm tree}| = 2\sqrt{2}\approx 2.83$.}
\label{fig04}
\end{figure}

In ensemble of matrices $A$ the percolation point is $q_c=1/N$ (the true phase transition is
defined in the thermodynamic limit, $N\to\infty$, only). For $q>q_c$, the graph ${\cal G}$ has a
giant component whose size is proportional to $N$, accompanied with numerous small components
(almost all of them are trees); for $q<q_c$, the graph ${\cal G}$ is composed of the collection of small components (almost all of them are again trees). For $q N\ll 1$ the isolated vertices
dominate, contributing to the trivial spectral density, $\rho(\lambda)=\delta(\lambda)$. The most
interesting region is close to $q_c$. In the scaling window $|q-q_c|=O(N^{-4/3})$, the size of the largest component scales as $N^{2/3}$. In what follows we discuss the behavior of the spectral
density $\rho(\lambda)$ for $q$ in this region. An example, in \fig{fig04}a we depict a collection of subgraphs for some randomly generated $N\times N = 500\times 500$ adjacency matrix at $q=1/N$ is
shown. Note that the components are predominantly linear subgraphs. In other realizations we
observed components with a single loop, but they are rare, usually a few per realization; more
complicated components, namely those with more than one loop, appear very rarely. In \fig{fig04}b
we show a spectral density, plotted as $\rho^{1/3}(\lambda)$, for an ensemble of 1000 random symmetric adjacency matrices $A$, each of size $N=500$ at $q=1/N$. Near the value $\lambda=\pm 2$ the enveloping curve exhibits a clearly visible fracturel. The choice of the exponent 1/3 for the curve $\rho^{1/3}(\lambda)$ is an "experimental" selection: we have just noted that such a scaling gives the best linear shape for the enveloping curve and clearly demonstrates the fracture of the shape near $\lambda=\pm 2$. The spectrum possesses the regular ultrametric hierarchical structure which is ultimately related to number-theoretic properties of some modular functions with discrete symmetries of the Lobachevsky plane.

Our discussion provides an example of a nontrivial manifestation of rare-event Bernoullian noise in some collective observables in large systems (like spectral density, $\rho(\lambda)$, for example).
From this point of view, we want to warn researchers working with effects of ultra-low doses
of chemicals: in the statistical analysis of, say, absorption spectra of extremely diluted solutions of chain-like polymer molecules, the valuable signal should be clearly purified from the background noise, which itself could have a very peculiar ultrametric shape.

\subsection{Phyllotaxis}
\label{sec14}

\begin{figure}[ht]
\centerline{\includegraphics[width=5cm]{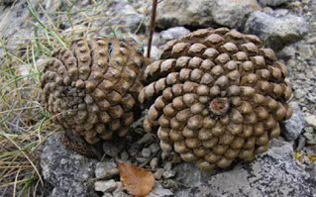} \hspace{5mm} \includegraphics[width=5cm]{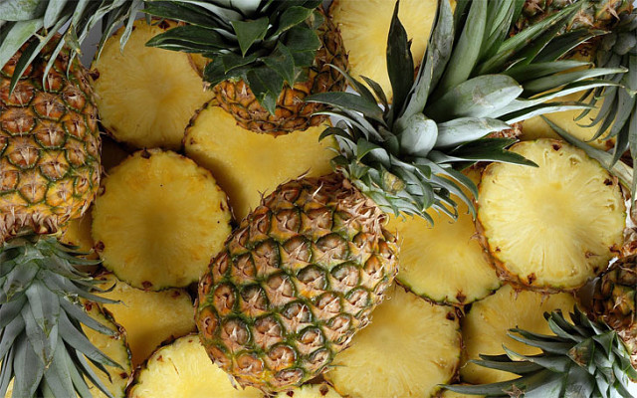}}
\caption{Cone (left), pineapple (right).}
\label{fig05}
\end{figure}

Amusing connection of cell packing with Fibonacci sequences, known as phyllotaxis, was observed
long time ago in studies of many naturalists, and currently is one of most famous manifestations of number theory in natural science -- see \fig{fig05}a,b. The generic description of growing plants based on symmetry arguments, allowed to uncover the role of Farey sequences in the plant's structure formation, however the question why nature selects just the Fibonacci sequence among other possible Farey ones, was hidden until modern time. A tantalizing answer to this question has been given by L. Levitov in 1990 (see Ref. [\refcite{levitov1}]) who proposed an "energetic" approach to the phyllotaxis, suggesting that the development of a plant is connected with an effective motion along the \emph{geodesics} on a Riemann surface associated with the energetic relief of a growing tissue.

The energetic mechanism suggested in Ref. [\refcite{levitov1}] was applied later in [\refcite{levitov3}] to the investigation of geometry of flux lattices pinned by layered superconductors. It has been shown that under the variation of magnetic field, the structure of the flux lattice can undergo a sequence of rearrangements encoded by the Farey numbers. However, dynamically accessible lattices are characterized by the subsequence of the Farey set, namely, by the Fibonacci numbers. Very illuminating experiments have been provided in Ref. [\refcite{audoly}] for lattice formed by drops in rotating liquid, and in Ref. [\refcite{cactus}] for equilibrium structure of "magnetic cacti".

\begin{figure}[ht]
\centerline{\includegraphics[width=10cm]{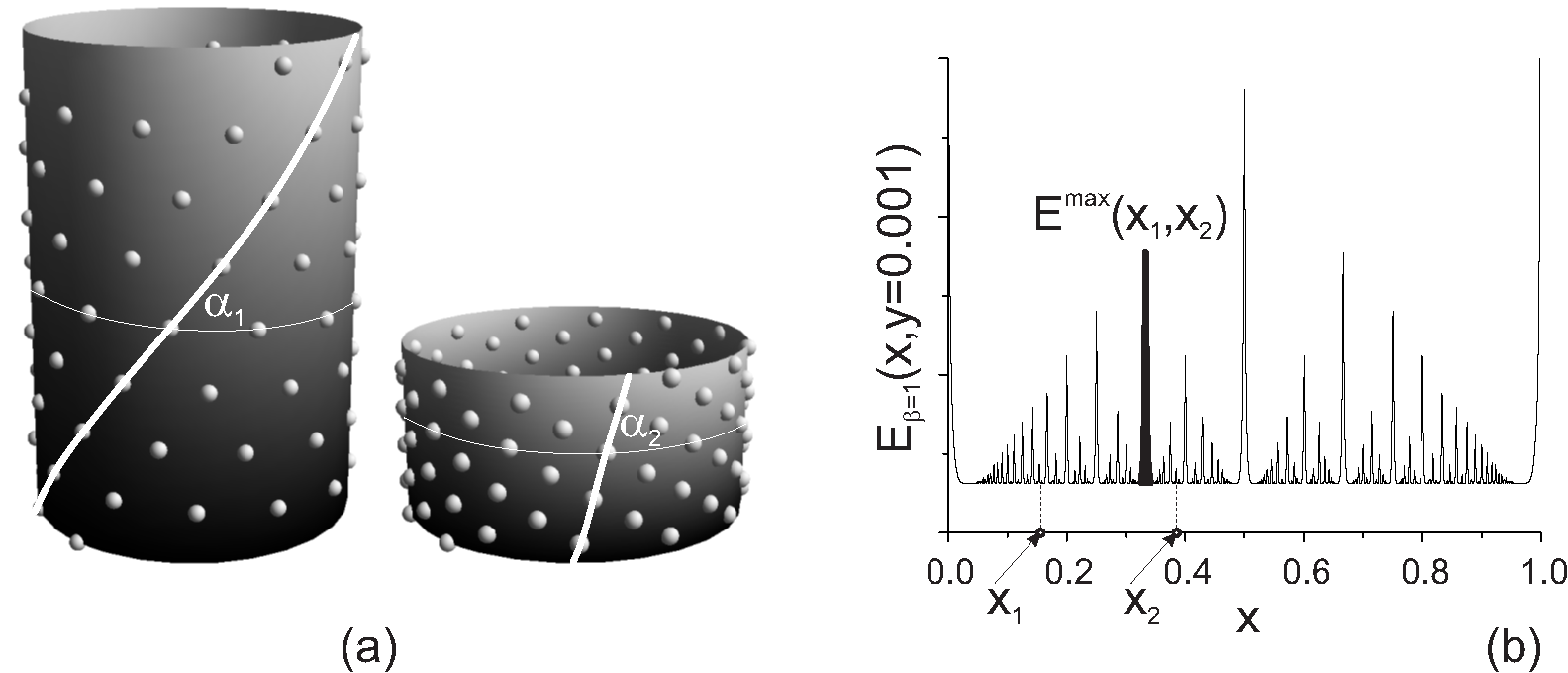}}
\caption{(a) Lattice of strongly repulsing particles on a surface of a cylinder under two distinct
compressions along the height for constant number of particles and radius of the cylinder; (b)
Dependence $E_{\beta}(x, y={\rm const})$ for strongly compressed lattice ($y=0.001<<1$ and
$\beta=1$).}
\label{fig06}
\end{figure}

Following L. Levitov, consider the model system of $N$ strongly repulsing particles equilibrated on the surface of a cylinder of fixed diameter, $D$, and hight, $H$, and look at the rearrangement of these particles when the cylinder is compressed along its height under the condition that $N$ and $D$ remain unchanged -- see \fig{fig06}a. At the continuous compression, for each height, the particles form a lattice with a minimal energy. Different lattice topologies, parameterized by the modular parameter, $\tau=D+iH$ represent in the phase space the valleys separated by energetic barriers having non-Archimedian ultrametric structure and the lattice ground state energy, $E(\tau)$, can be expressed in terms of the so-called modular Dedekind $\eta$-function. Moreover, one can relate $E(\tau)$ to the spectral density, $\rho(\tau)$, of an ensemble of exponentially distributed bi-diagonal operators with the off-diagonal disorder. The ultrametric organization of the energy relief, $E_{\beta}(x,y={\rm const})$ as a function of $x$, should be understood as follows. Identifying the energy with the metrics, namely setting $d(x_1,x_2)=E^{\rm max}(x_1,x_2)$ as shown in the \fig{fig06}b we immediately arrive at the picture which resembles the \fig{fig04}. In details this scenario is considered in the Section \ref{sec3}.

In Secton \ref{sec3} we claim that the ultrametricity is the consiquence of number-theoretic collective properties of distributions appearing in discrete physical systems when some observable is a quotient of two independent exponentially weighted integers. The example is the density of eigenvalues of the ensemble of linear polymer chains distributed with the exponential law $\sim f^L$ ($0<f<1$), where $L$ is the chain length. As $f\to 1$, the spectral density can be expressed through the function, discontinuous at all rational points, known as Thomae ("popcorn") function. We suggest a continuous approximation of the popcorn function, based on the Dedekind $\eta$-function near the real axis. Moreover, we provide simple arguments, based on the
"Euclid orchard" construction, that demonstrate the presence of Lifshitz tails at the spectral edges, typical for the 1D Anderson localization. We emphasize that the ultrametric structure of the spectral density is ultimately connected with number-theoretic properties of modular functions.

\section{Buckling of thin tissues and hyperbolic geometry}
\label{sec2}

\subsection{Basic facts about the eikonal equation}
\label{sec21}

To make the content of the paper as self-contained as possible, it seems instructive to provide
some important definitions used at length of the paper. The key ingredient of our consideration is
the "eikonal" equation, which is the analogue of the Hamilton-Jacobi equation in geometric
optics. As we show below, the eikonal equation provides optimal embedding of an exponentially
growing surface into the 3D Euclidean plane. Meaning of the notion "optimal" has two different
connotations in our approach:
\begin{itemize}
\item[i)] On one hand, from viewpoint of the Hamilton-Jacobi theory, the eikonal equation appears in the minimization of the action $A=\int_\gamma L dt$ with some Lagrangian $L$. According to the Fermat principle, the time of the ray propagation in the inhomogeneous media with the space-dependent refraction coefficient, $n(x,y)$, should be minimal.
\item[ii)] On the other hand, the eikonal equation emerges in our work in a purely geometric setting following directly from the conformal approach.
\end{itemize}

First attempts to formulate classical mechanics problems in geometric optics terms goes back to the works of Klein [\refcite{klein}] in 19th century. His ideas contributed to the corpuscular theory in a short-wavelength regime, as long as the same mechanical formalism applied to massless particles,was consistent with the wave approach. Later, in the context of general relativity, this approach was renewed to treat gravitational field as an optic medium [\refcite{rumer}].

The Fermat principle states that the time $dt$ for a ray to propagate along a curve $\gamma$
between two closely located points $M({\bf x})$ and $N({\bf x}+d{\bf x})$ in an inhomogeneous
media, should be minimal. The total time $T$ can be written in the form $T=\frac{1}{c}\int_{M}^{N} n(\mathbf{x}(s))ds$ where $n(\mathbf{x})=\frac{c}{\textrm{v}(\mathbf{x})}$ is the refraction
coefficient at the point $\mathbf{x}=\{x^i\}$ of a $D$-dimensional space $(i=1,...,D)$, $c$ and
$\textrm{v}(\mathbf{x})$ are correspondingly the light speeds in vacuum and in the media, and
$d|\mathbf{x}|=ds$ is the spatial increment along the ray. Following the optical-mechanical
analogy, according to which the action in mechanics corresponds to eikonal in optics, one can write down the "optic length" or eikonal, $S=cT$ in Lagrangian terms: $S=\int_{M}^{N} L(\mathbf{x},
\dot{\mathbf{x}}) ds$ with the Lagrangian $L(\mathbf{x},\dot{\mathbf{x}})= n(\mathbf{x}(s))
\sqrt{\dot{\mathbf{x}}(s)\dot{\mathbf{x}}(s)}$, where $\dot{\mathbf{x}}^2= \sum_{i=1}^D
\big(\frac{dx^i}{ds}\big)^2$. We would like to mention here, that optical properties of the media
can be also treated in terms of induced Riemann metrics in vacuum:
\be
S=\int_{M}^{N}n(\mathbf{x}(s)) ds = \int_{M}^{N} \sqrt{\dot{\mathbf{x}} g(\mathbf{x})
\dot{\mathbf{x}}} ds
\label{eq:eikonal-metrics}
\ee
where $g_{ij} = n^2(\mathbf{x})\delta_{ij}$ stands for induced metrics components in isotropic media case. Thus, from the geometrical point, the ray trajectory can be understood as a "minimal curve" in a certain Riemann space. This representation suggests to consider optimal ray paths as geodesics in the space with known metrics $g$.

Stationarity of optic length, $S$, i.e. $\delta S=0$, together with the condition
$|\dot{\mathbf{x}}|=1$, defines the Euler equation:
\be
\frac{d}{ds}\left(n(\mathbf{x})\frac{d\mathbf{x}}{ds}\right)= \nabla n(\mathbf{x})
\label{eq:euler}
\ee
from which one can directly proceed to the Huygens principle by integrating \eq{eq:euler} over $s$:
$\nabla S(\mathbf{x})=n(\mathbf{x})\frac{d\mathbf{x}}{ds}$. Squaring both sides of the latter
equation we end up with the eikonal equation:
\be
\left(\nabla S(\mathbf{x})\right)^2=n^2(\mathbf{x})
\label{eq:eikonal}
\ee
The eikonal equation Eq.\eq{eq:eikonal} has the same form as the Hamilton-Jacobi equation in
mechanics for action in the $D+1$-dimensional space, which in turn can be understood as the
relativistic equation for the light, propagating in the Riemannian space.

\subsection{The model: formalization of physical ideas}
\label{sec22}

\begin{figure}[ht]
\centering
\includegraphics[width=8cm]{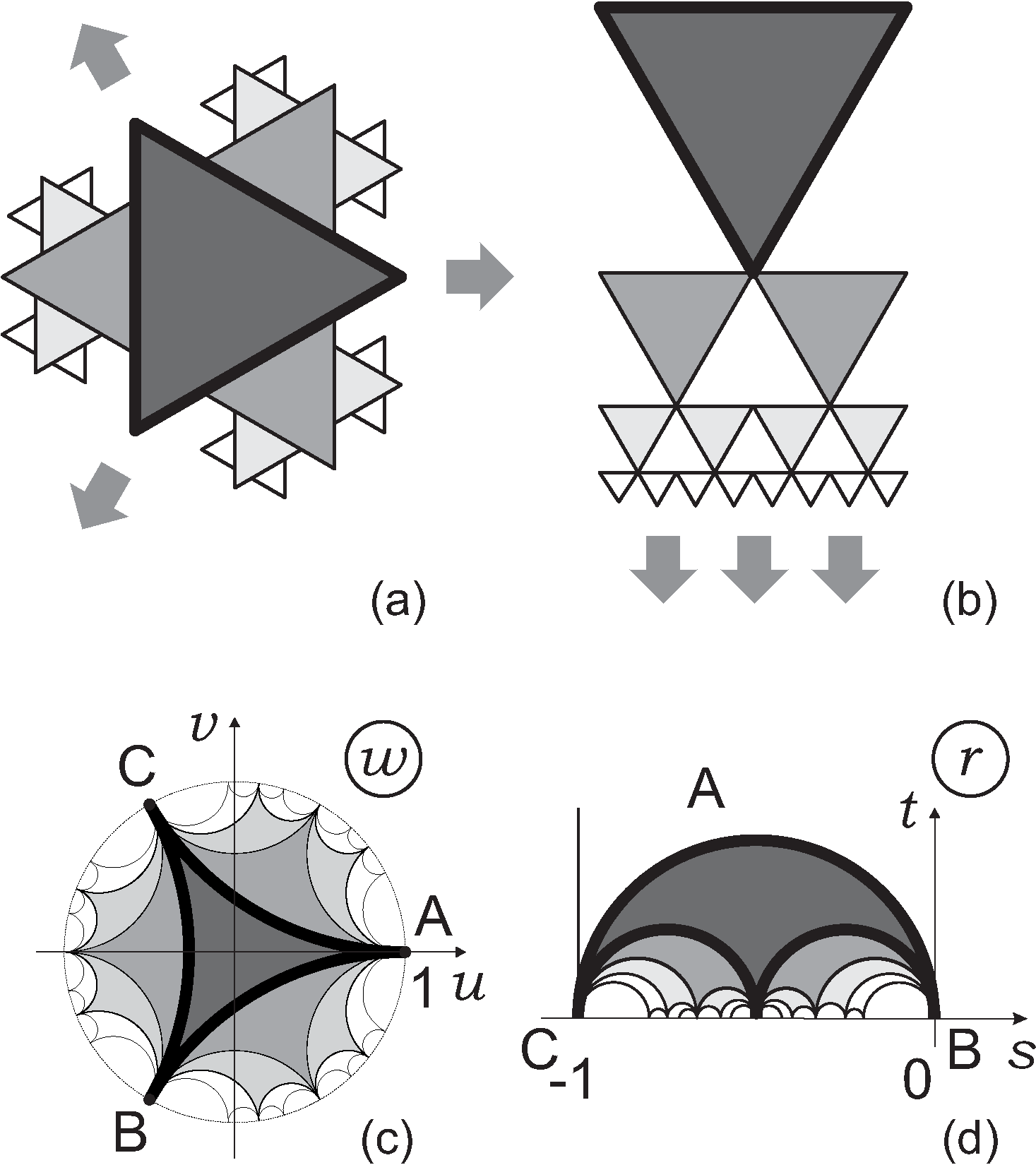}
\caption{(a) Uniform two-dimensional hyperbolic growth out of the unit domain in the plane;
(b) One-dimensional hyperbolic growth out of the linear segment; c) Tessellation of the hyperbolic
Poincar\'e disc by the images of flat Euclidean triangles; d) Tessellation of the domain in the
hyperbolic half-plane by the images of flat Euclidean triangles.}
\label{fig07}
\end{figure}

We discuss buckling phenomena for two different growth symmetries shown schematically in the
\fig{fig07}a-b:
\begin{itemize}
\item[i)] uniform two-dimensional division from the point-like source (\fig{fig07}a),
\item[ii)] directed one-dimensional growth from the linear segment (\fig{fig07}b). In \fig{fig07}a-b different generations of cells are shown by the shades of gray.
\end{itemize}

For convenience of perception, sizes of cells in each new generation are decreasing in geometric progression, otherwise it would be impossible to draw them in a 2D flat sheet of paper and the figure would be incomprehensible. In Figs. \ref{fig07}c,d we imitate the protocols of growth depicted above in Figs. \ref{fig07}a,b by embedding the exponentially growing structure in the corresponding plane domain equipped with the hyperbolic metrics. The advantage of such embedding consists in the possibility to continue all functions smoothly through the boundaries of elementary domains, that cover the whole plane without gaps and intersections. Details of this construction and its connection to the growth in the 3D Euclidean space are explained below.

It is known (see, for example Ref. [\refcite{marder}]) that the optimal buckling surface is fully determined by the metric tensor through the minimization of the discrete functional of special energetic form. Namely, define the energy of a deformed thin membrane, having buckling profile $f(x,y)$ above the domain, parameterized by $(x,y)$, as:
\be
E\{f(x,y)\} \sim \sum_{i,j} \left(\Big(f_{ij}\Big)^2-\sum_{\alpha,\beta}\Delta^{\alpha}_{ij}
g_{\alpha\beta} \Delta^{\beta}_{ij} \right)^2
\label{eq:E}
\ee
where $g_{\alpha\beta}$ is the induced metrics of the membrane, $f_{ij} \equiv |f(x_i, y_i) -
f(x_j, y_j)|$ is the distance between neighboring points and $\Delta_{ij}$ is the equilibrium
distance between them. The typical (optimal) shape $\bar{f}(x,y)$ is obtained by minimization of
\eq{eq:E} for any rigidity. However, the metric tensor, $g_{\alpha\beta}$ is a priori unknown
since its elements depend on specifics of the differential growth protocol, therefore some
plausible conjectures concerning its structure should be suggested. For example, in Ref. [\refcite{marder}] a directed growth of a tissue with one non-Euclidean metrics component, $g_{xx}(y)$, was considered. The diagonal component $g_{xx}(y)$ was supposed to increase exponentially in the direction of the growth, $y$, and crumpling of a leaf near its edge was finally established and analyzed.

\subsection{Conformal approach to buckling}
\label{sec23}

The preset rules of uniform exponential cells division determine the structure of the hyperbolic
graph, $\gamma$, while the infinitesimal membrane thickness allows for the isometrical embedding of the graph $\gamma$ into the 3D space. We exploit conformal and metric relations between the surface
structure in the 3D space and the graph $\gamma$ embedded into the flat domain with the hyperbolic metrics. The embedding procedure consists of a sequence of conformal transformations with a
constraint on area preservation of an elementary plaquette. This eventually yields the knowledge of the Jacobian (the "coefficient of deformation"), $J(x,y)$, for the hyperbolic surface, which is
embedded into the 3D space via the orthogonal projection. Equipped by the key assumption, that a
smooth yet unknown surface $f(x,y)$ is \emph{function}, our procedure straightforwardly implies a
differential equation on the optimal surface. Note that a version of the Amsler surface cannot be
reconstructed in the same way since it is not a function above some planar domain.

\begin{figure}[ht]
\centering
\includegraphics[width=10cm]{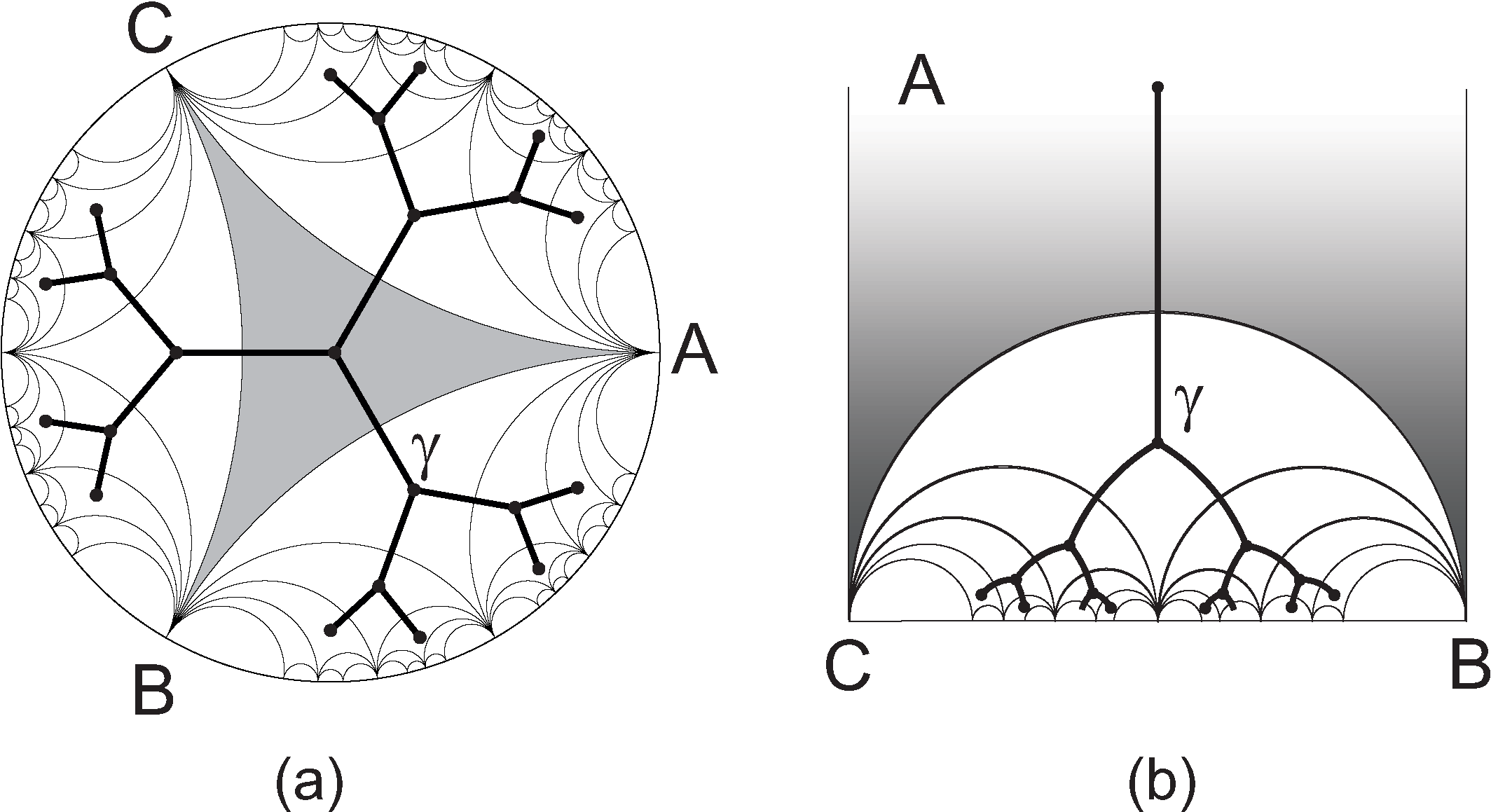}
\caption{Tessellation of the hyperbolic plane by the images of the
curvilinear triangle $ABC$: (a) for Poincar\'e disc; (b) for a strip of the upper half-plane. The
graph $\gamma$ connects the centers of images of $ABC$.}
\label{fig08}
\end{figure}

To realize our construction explicitly, we embed isometrically the graph $\gamma$ (as it has been mentioned above):
\begin{itemize}
\item[i)] into the Poincar\'e disk ($|w|<1$) for the model of uniform planar growth,
\item[ii)] into the strip of the half-plane ($\im r > 0, -1<\re r<0$) for the model of one-dimensional growth.
\end{itemize}
In the \fig{fig08} we have drawn the tessellation of the Poincar\'e disc and of the strip by equilateral curvilinear triangles, which are obtained from the flat triangle $ABC$ of the hyperbolic surface by conformal mappings $z(w)$ and $z(r)$ discussed below (see Ref. [\refcite{polovnikov}] for details). Note, that a conformal mapping preserves the angles between adjacent branches of the graph. The graph $\gamma$, shown in the \fig{fig08}, connects the centers of the triangles and is isometrically embedded into the corresponding hyperbolic domain. Besides, the areas of images of the domain ABC are the same.

For the sake of definiteness consider the graph $\gamma$, isometrically embedded into the
hyperbolic disk, shown in the \fig{fig08}a. Now, we would like to find the surface in the 3D
Euclidean space above the $w$-plane such that its Euclidean metrics coincides with the
non-Euclidean metrics in the disk. The Hilbert theorem [\refcite{hilbert}] prohibits to do that for the class of $C^2$-smooth surfaces. However, since we are interested in the isometric embedding of piecewise surface consisting of glued triangles of fixed area, we can proceed with the standard
arguments of differential geometry [\refcite{diff-geom}]. The metrics $ds^2$ of a 2D surface,
parameterized by $(u, v)$, is given by the coefficients
\be
E=\mathbf{r}_u^2, \quad F=\mathbf{r}_v^2, \quad G=(\mathbf{r}_u, \mathbf{r}_v)
\ee
of the first quadratic form of this surface:
\be
ds^2=E\, du^2 + 2F\, du dv + G\, dv^2
\ee
The surface area then reads $dS= \sqrt{EG-F^2}\, du dv$.

The area $S_{ABC}$ of the planar triangle $ABC$ on the plane $z = x + iy$ can be written as:
\be
S_{ABC}=\int\limits_{\triangle ABC} dx dy = \mathrm{const}
\label{eq:03}
\ee
where the integration is restricted by the boundary of the triangle. Since we aimed to conserve the metrics, let us require that the area of the hyperbolic triangle $ABC$, after the conformal
mapping, is not changed and, therefore, it reads:
\be
S_{ABC} = \int\limits_{\triangle ABC} |J(z,w)| du dv; \quad J(z,w)=\left|\begin{array}{cc} \disp
\partial_u x & \disp \partial_u y \medskip  \\ \disp
\partial_v x & \disp \partial_v y
\end{array}\right|
\label{eq:04}
\ee
where $J(z,w)$ is the Jacobian of transition form $z$ to new coordinates, $w$. If $z(w)$ is
holomorphic function, the Cauchy-Riemann conditions allow to write
\be
J(w) = \left|\frac{dz(w)}{dw}\right|^2\equiv |z'(w)|^2.
\ee
From the other hand, we may treat the value of the Jacobian, $J(w)$, as a factor relating the
change of the surface element under transition to a new metrics, the co-called "coefficient of
deformation". As long as the metrics in the hyperbolic domain should reproduce the Euclidean metrics of the smooth surface, $f(u,v)$, one should set $J = \sqrt{EG-F^2}$, where $E,G,F$ are the coefficients of the first quadratic form of the surface $f$. Now, if $f(u,v)$ is function above $w$-plane, its Jacobian adopts a simple form:
\be
J(u,v) = \sqrt{1+(\partial_u f)^2+(\partial_v f)^2}
\label{eq:06}
\ee

Making use of the polar coordinates in our complex $w$-domain, $\{(\rho, \phi): u = \rho\cos\phi, v = \rho\sin\phi\}$, we eventually arrive at nonlinear partial differential equation for surface profile $f(\rho,\phi)$ in the polar coordinates above $w$:
\be
\Big(\partial_\rho f(\rho,\phi)\Big)^2+\frac{1}{\rho^2}\Big(\partial_\phi
f(\rho,\phi)\Big)^2=|z'(w)|^4-1
\label{eq:08}
\ee

In the case of the hyperbolic strip domain, \fig{fig08}b, the equation for the growth profile above the domain can be written in local cartesian coordinates, $r = s + it$:
\be
\Big(\partial_s f(s, t)\Big)^2+\Big(\partial_t f(s,t)\Big)^2=|z'(r)|^4-1
\label{eq:07}
\ee

Note, that the inequalities $|z'(w)| > 1, |z'(r)| > 1$, following from \eq{eq:08}-\eq{eq:07}, determine the local condition of existence of non-zero real solution and, as we discuss below, can be interpreted as the presence of a finite scale surface rigidity.

To establish a bridge between optic and growth problems, let us mention that, say,
equation \eq{eq:08}, coincides with the two-dimensional eikonal equation \eq{eq:eikonal} for the
wavefront, $S(w)$, describing the light propagating according the Huygens principle in the
unit disk with the refraction coefficient
\be
n(w) = \sqrt{|z'(w)|^4-1}
\label{eq:n}
\ee

\begin{figure}[ht]
\centering
\includegraphics[width=7cm]{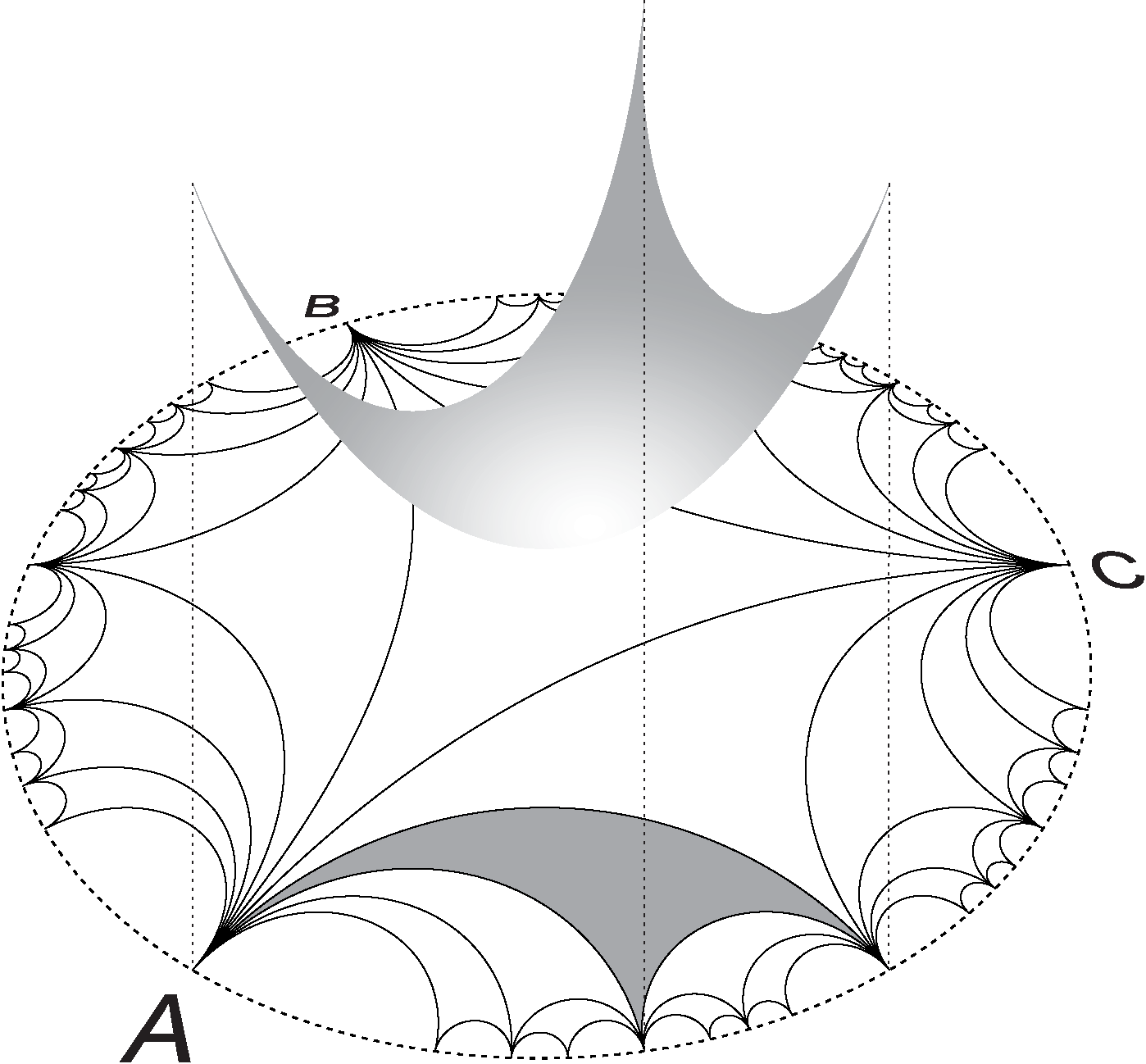}
\caption{Orthogonal projection above the Poincar\'e disc: area of the curvilinear triangle in Euclidean space coincides with the area of the triangle in hyperbolic metrics in Poincar\'e disc.}
\label{fig09}
\end{figure}

We can construct the conformal mappings $z(r)$ and $z(w)$ of the flat equilateral triangle $ABC$ in the Euclidean complex plane $z=x+iy$ onto the circular triangle $ABC$ in the complex domains $r = s
+ it$ and $w=\rho(\cos\phi+i\sin\phi)$ correspondingly. The absolute value of the Gaussian
curvature is controlled by the number, $V$, of equilateral triangles glued in one vertex: the
surface is hyperbolic only for $V>6$. The surfaces with any $V>6$ have qualitatively similar
behavior, however the simplest case for analytical treatment corresponds to $V=\infty$, when the
dual graph $\gamma$ is loopless. The details of the conformal mapping of the flat triangle with
side $a$ to the triangle with angles $\{0,0,0\}$ in the unit strip $r$ are given in Ref. [\refcite{polovnikov}]. The Jacobian $J(z(r))$ of conformal mapping $z \to r$ reads:
\be
J(r)=|z'(r)|^2 =\frac{h^2}{a^2}|\eta(r)|^8; \quad \eta(w)=e^{\pi i w/12}\prod_{n=0}^{\infty}(1-e^{2\pi i n w})
\label{eq:ded1}
\ee
and the Jacobian of the mapping $z \to w$, is written through the function $r(w)$ that conformally
maps the triangle from the strip onto the Pincar\'e disk:
\be
J(w)=|z'(w)|^2 =\frac{3h^2}{a^2} \frac{|\eta(r(w))|^8}{|1-w|^4}
\label{eq:ded2}
\ee
where
\be
r(w) = e^{-i\pi/3}\frac{e^{2i\pi/3}-w}{1-w}-1; \quad h= \left(\frac{16}{\pi}\right)^{1/3}
\frac{\Gamma(\frac{2}{3})}{\Gamma^2(\frac{1}{3})} \approx 0.325
\ee
In \eq{eq:ded1} and \eq{eq:ded2}, the function $\eta(...)$ is the so-called Dedekind $\eta$-function [\refcite{jacobi}].

\subsection{Physical interpretation of results and conjectures}
\label{sec24}

The eikonal equation, \eq{eq:eikonal}, with \emph{constant} refraction index, $n$, corresponds to
optically homogeneous 2D domain, in which the light propagates along straight lines in Euclidean
metrics. From the other hand, in this case the eikonal equation yields the action surface with zero Gaussian curvature: a conical surface above the disk, $S(\rho, \phi) \sim \rho$, for the uniform 2D
growth and a plane above the strip, $S(s, t) \sim t$, for the directed growth. Note, that at least one family of geodesics of these surfaces consists of lines that are projected to the light
propagation paths in the underlying domain. We will show below that the geodesics of the eikonal
surface conserve this property even when the media becomes optically inhomogeneous.

For growth, the constant refraction index corresponds to an isometry of a planar growing surface
and absence of buckling. The conformal transformation, that results in the corresponding "coefficient of deformation", $J^2(u, v) = n^2(u, v) + 1$, is uniformly compressive and the tissue remains everywhere flat. Thus, it becomes clear, why the essential condition for buckling to appear is the \emph{differential growth}, i.e. the spatial dependence of local rules of cells division.

We solve \eq{eq:08} and \eq{eq:07} numerically with the Jacobian, corresponding to exponentially
growing circumference, \eq{eq:ded1}-\eq{eq:ded2}, for different parameters $a$. We have chosen the Dirichlet initial conditions along the line (for directed growth above the strip) and along the
circle of some small enough radius (for uniform 2D growth above the disk). The right-hand side of the eikonal equations for the specific growth protocol is smooth and nearly constant up some radius and then becomes more and more rugged. The constant plateau in vicinity of initial stages of growth is related with the fact, that exponentially dividing cells can be organized in a Euclidean plane up to some finite generations of growth. However, as the cells proliferate further, the isometry of
their mutual disposition becomes incompatible with the Euclidean geometry and buckling of the
tissue is observed. Note, that the Jacobian is angular-dependent, that is the artefact of chosen
triangular symmetry for the cells in our model. The existence of real solution, $\bar f(u,v)$ of
the eikonal equation is related to the sign of its right-hand side and is controlled by the parameter $a$, while the complex solution $f(u, v) = f_R(u, v) + if_I(u, v)$ can be found for every $a$.

\begin{figure}[ht]
\centering
\includegraphics[width=10cm]{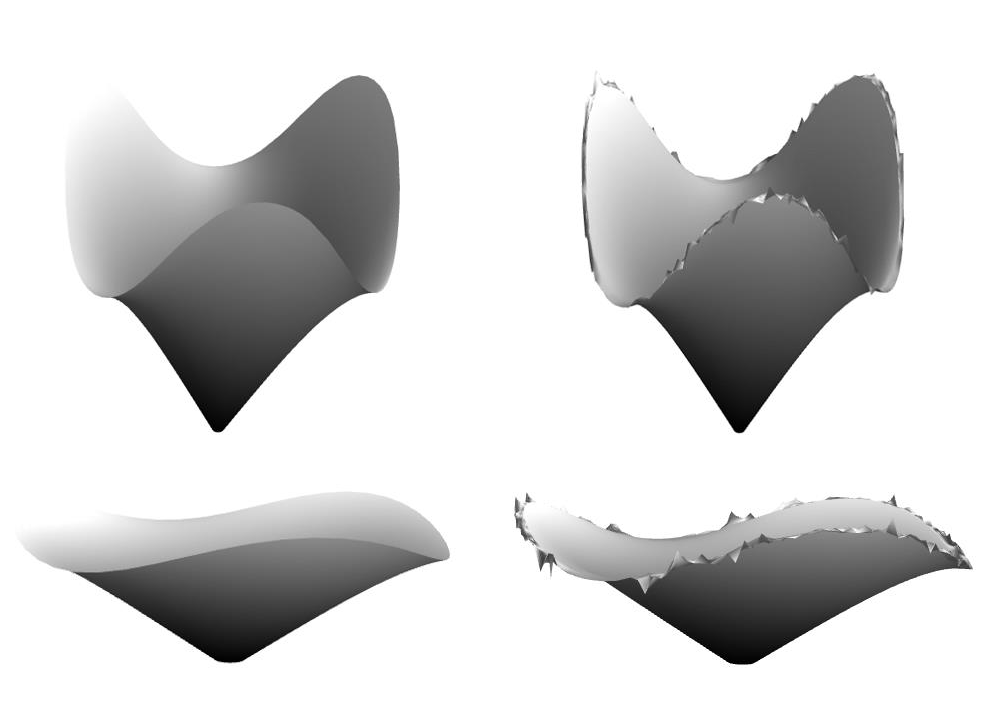}
\caption{The bluebell phase, obtained by numerical solution of \eq{eq:08} for two flexible tissues:
$a = 0.07$ (first row) and $a = 0.14$ (second row). Figures on the right show appearance of buckling instabilities at the edge with growth.}
\label{fig10}
\end{figure}

First, we consider the 2D growth above the Poincar\'e domain, starting our numerics from low
enough values of $a$, for which the right-hand side of the eikonal equation, \eq{eq:08}, is strictly positive on the plateau around the source of growth. Physically that means flexible enough tissues, since, by construction, we require $a$ to be a scale on which the triangulated tissue does not violate flat geometry. The real solution $\bar f(u,v)$ for these parameters exists up to late stages of growth, see \fig{fig10} left. Note, that a conical solution at early stages of growth is related to the plateau in the Jacobain and, as it was discussed above, corresponds to the regime when cells can find place on the surface without violation flat geometry. From the geometric optics point of view, that corresponds to constant refraction index and straight Fermat geodesic paths in the underlying 2D domain. We show in the \fig{fig10} that under increasing of $a$ the initial area of conical behavior is shrinking, since the critical generation, at which the first buckling mode appears, is lower for larger cells. In course of growth, the surface is getting negatively curved for some angular directions, consistent with chosen triangular symmetry. It is found reminiscent of the shape of bluebells and, in general, many sorts of flowers.

At late stages of growth, as we approach the boundary of the Poincar\'e disk, $\rho \to 1$ at some fixed value of $\phi$, corresponding values of the right hand side of \eq{eq:08} become negative,
leading to the complex solution of the eikonal equation. Fortunately, we may infer some useful
information from the holomorphic properties of the eikonal equation in this regime, not too close
to the boundary of the disc. Applying the Cauchy-Riemann conditions to the solution of the eikonal equation, $f$, we have: $\partial_u f_R = \partial_v f_I$ and $\partial_v f_R = -\partial_u f_I$.
Thus, the function $\bar f$ can be analytically continued in the vicinity of points along the curve $\Gamma$ in the $(uv)$ plane, at which the right hand side of the eikonal equation nullifies. Moreover, using this property, one can show, that the absolute value of the complex solution in the vicinity of $\Gamma$ smoothly transfers to the real-valued solution, as one approaches the $\Gamma$ curve:
\be
\begin{array}{rll}
\disp \lim_{(u, v) \to \Gamma} \left(\nabla |f(u, v)|\right)^2 &=& \left(\nabla f_R(u,
v)\right)^2|_{\Gamma} \equiv \left(\nabla \bar f(u, v) \right)^2 \medskip \\
\disp |f(u, v)| &=& \sqrt{f_R^2(u, v) + f_I^2(u, v)}
\end{array}
\label{mod}
\ee

\begin{figure}[ht]
\centering
\includegraphics[width=10cm]{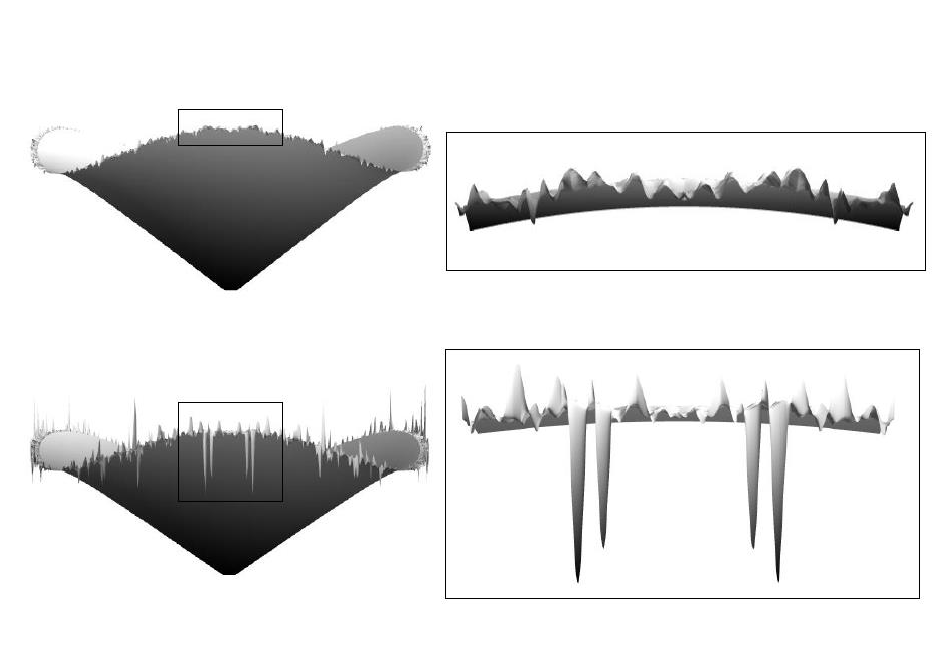}
\caption{Development of buckling instabilities at the edge of flower for rigidity parameter $a = 0.14$. The right figures show hierarchical organization of the flower's circumference in detail.}
\label{fig11}
\end{figure}

The non-existence of real solutions of the eikonal equation at late stages is a direct consequence of the presence of finite bending scale, on which the tissue is locally flat. As it was
mentioned above and is shown in the \fig{fig10}, low values of $a$ lead to elongated conical
regime. Since $a$ stands for the scale on which the circumference length of the tissue doubles, in the $a \to 0$ limit the real solution exists everywhere inside the disk, but it is everywhere flat (conical). Hopefully, the analytic continuation allows one to investigate buckling for negative values of $n^2(u,v) = J^2(u, v)-1$ by taking the absolute value of the solution, at least not far away from the zero-curve $\Gamma$. In this regime buckling instabilities on the circumference of the bluebell arise. In the \fig{fig11} we show proliferation of buckling near the critical point. First, the evolution of buckling instabilities at the edge can be understood as a subsequent doubling of peaks and saddles along the direction of growth. Then some hierarchy in peaks size is seen. We note, that this hierarchical organization is a natural result due to the theoretic-number properties of the Dedekind $\eta$-function. Though it is known, that in real plants and flowers buckling instabilities do not proliferate profoundly, since the division process is getting limited at late stages of growth, the formal continuation of the eikonal equation beyond $\Gamma$ predicts a self-similar buckling profile at the circumference of growing tissues.

\begin{figure}[ht]
\centering
\includegraphics[width=10cm]{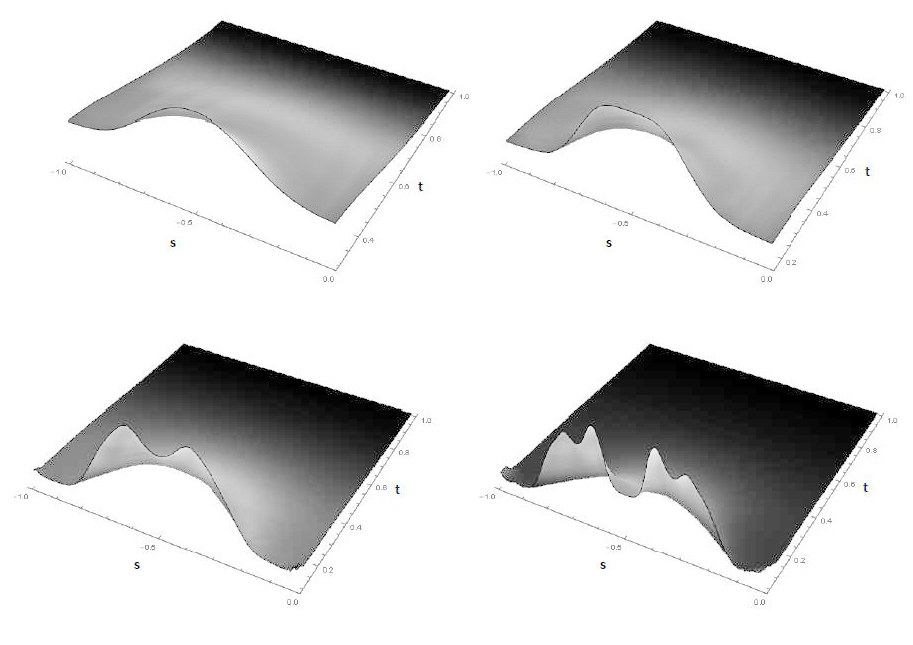}
\caption{Numerical solutions of \eq{eq:07} for the directed growth.
Figures show enhancing of buckling at the edge.}
\label{fig12}
\end{figure}

Now we pay attention to the directed growth above the half-plane domain. Here we solve the equation \eq{eq:07} with the Dirichlet boundary conditions, set along the line $t = 1$, and the tissue is
growing towards the boundary $t = 0$ in the upper halfplane $\im r>0$. At low stages of growth
the solution is flat until the first buckling mode appear, \fig{fig12}. The subsequent
growth is described by taking the absolute value of the solution, since no real solution exists
anymore. As in the former case, the behavior is controlled by the value of $a$.

When the growth approaches the boundary, the edge of the tissue becomes more and more wrinkled.
Emergence of new buckling modes is the consequence of the Dedekind $\eta$-function properties: doubling of parental peaks at the course of growth. Under the energetic approach for a leaf, very similar fractal structures can be inferred from the interplay between stretching and bending energies in the limit of extremely thin membranes: while the cell density (and the corresponding strain, $\sigma$) on the periphery increases, the newly generating wavelengths decrease, $\lambda\sim \sigma^{-1/4}$ as discussed in Ref. [\refcite{mahadevan}].

Increasing the size $a$ of the elementary flat triangle domain, we figure out, that for some
critical value, $a_{cr}$, the starting plateau of the corresponding Jacobian crosses the zero level and becomes negative. Our model implies no solutions for such stiff tissues. This limitation is
quite natural since we do not consider in-plane deformations of the tissue. In reality, for $a
> a_{cr}$ the tissue is so stiff, that it turns beneficial to be squeezed in-plane rather than to
buckle out. One may conjecture that $a$ is the analogue of the Young modulus, $E$, that is
known to regulate the rigidity of the tissue in the energetic approach, along with the thickness,
$h$, and the Poisson modulus, $\mu$, in their certain combination, known as bending stiffness, $D =
\frac{Eh^3}{12(1-\mu^2)}$.

\begin{figure}[ht]
\centering
\includegraphics[width=10cm]{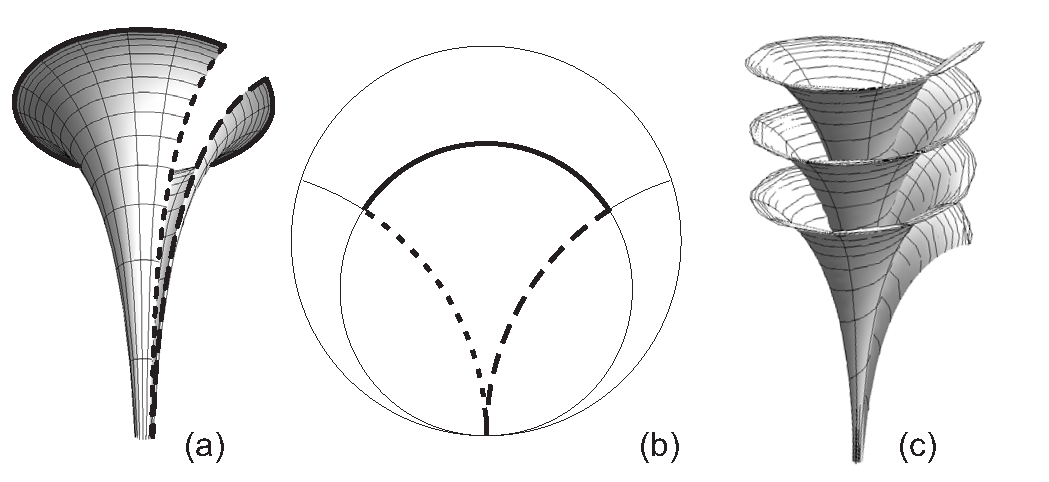}
\caption{(a)-(b) Pseudosphere and correspondence of boundaries on the Poincar\'e disc; (c) Dini
surface.}
\label{fig13}
\end{figure}

It is worth mentioning that at first stages of growth, until the instabilities at the circumference have not yet appeared, at certain angles (triangle-like cells) the surface bends similar to the Beltrami's pseudosphere, that has a constant negative curvature at every point of the surface (compare \fig{fig10} and \fig{fig13}). The similarity is even more striking for very low $a$, when the triangulating parameter is fairly small. It is known that the pseudosphere locally realizes the Lobachevsky geometry and can be isometrically mapped onto \emph{the finite part} of the half-plane or of the Poincar\'e disk, \fig{fig13}a-b. We have mentioned already that according to the Hilbert theorem, [\refcite{hilbert}], no full isometric embedding of the Poincar\'e disk into the 3D space exists. Thus, in order to organize itself in the 3D space, the plant grows by the cascades of pseudospheres, resembling peaks and saddles, that is an alternative view on essence of buckling.

Interestingly, some flowers, such as calla lilies, initially grow psuedospherically, but then crack at some stage of growth and start twisting around in a helix. Apparently, this is another route of dynamic organization of non-Euclidean isometry in the Euclidean space. The Dini's surface, \fig{fig13}c is known in differential geometry as a surface of constant negative curvature and, in comparison with the Beltrami's pseudosphere, is infinite. The problem of sudden cracking of the lilies seems to be purely biological, but as soon as the crack appeared, the flower may relief the stresses caused by subsequent differential growth through twisting its petals in the Dini's fashion.

Turn now to the eikonal interpretation of buckling. For the sake of simplicity, we will proceed here in the cartesian coordinates. Seeking the solution of \eq{eq:eikonal} and \eq{eq:07} in the implicit form $H(\mathbf{x})\equiv H(x^0,x^1,x^2) = H^0$ with $x^0= i f$, $x^1=u$, $x^2=v$, we can rewrite \eq{eq:07} as:
\be
g^{ij}\frac{\partial H(\mathbf{x})}{\partial x^i}\frac{\partial H(\mathbf{x})}{\partial x^j}= 0;
\quad g^{ik} = \left(\begin{matrix} n^2 & 0 & 0 \\ 0 & 1 & 0 \\ 0 & 0 & 1 \end{matrix}\right)
\label{eq:rel}
\ee
Eq.\eq{eq:rel} reveals the relativistic nature of the eikonal equation [\refcite{rumer}] and describes the propagation of light in a (2+1)D space-time in the gravitational field with induced metrics $g$ defined by the metric tensor $g^{ik}$, where $n\equiv n(x^1,x^2)$, speed of light put $c=1$. Having $g$, one can reconstruct geodesics that define the paths of the light propagation in our space-time. The parameterized geodesics family, $x^{\lambda}(\tau)$, where $\lambda = 0, 1, 2$, can be found from the equation:
\be
\frac{d^2 x^{\lambda}}{d \tau^2} + \Gamma^{\lambda}_{ij} \frac{dx^i}{d \tau} \frac{dx^j}{d \tau}=0
\label{geo}
\ee
where
\be
\Gamma^{i}_{kl} = \frac{1}{2} g^{im} \left(\frac{\partial g_{mk}}{\partial x^l} + \frac{\partial
g_{ml}}{\partial x^k} - \frac{\partial g_{kl}}{\partial x^m} \right)
\ee
are Christoffel symbols and $g_{ij}$ is the covariant form of the metrics ($g_{ij}g^{jk} =
\delta^k_i$). Calculating the symbols for the specific metrics \eq{eq:rel}, we end up with the set of equations for the geodesics in a parametric form:
\be
\left\{\begin{array}{l}
\disp u_{\tau \tau} - \frac{1}{n^3} \frac{\partial n}{\partial u} f_{\tau}^2 = 0, \medskip \\
\disp v_{\tau \tau} - \frac{1}{n^3} \frac{\partial n}{\partial v} f_{\tau}^2 = 0, \medskip \\
\disp f_{\tau \tau} - \frac{2}{n}\frac{d n}{d \tau} f_{\tau} = 0
\end{array} \right.
\label{traj}
\ee
From the first two lines of \eq{traj}, one gets $\frac{u_{\tau \tau}}{v_{\tau \tau}} =
\frac{\partial n}{\partial u} \left(\frac{\partial n}{\partial v}\right)^{-1}$. Note, that the same relation follows directly from \eq{eq:euler}, if the planar domain is parameterized by the same coordinates $\mathbf x = \mathbf x(u, v)$. Thus, one may conclude, that the projections of the geodesics from the (2+1)D space-time onto the $(uv)$-plane coincide with light trajectories in the flat domain with refraction coefficient $n(u,v)$.

In this paper we discussed the optimal buckling profile formation of growing two-dimensional tissue evoked by the exponential cell division from the point-like source and from the linear segment.
Such processes imply excess material generation enforcing the tissue to wrinkle as it approaches
the domain boundary. Resulting optimal hyperbolic surface is described by the eikonal equation for the two-dimensional profile, and allows for simple geometric optics analogy. It is shown that the
surface height above the domain mimics the eikonal (action) surface of a particle moving in the 2D media with certain refraction index, $n$, which, in turn, is linked to microscopic rules of elementary cell division and symmetry of the plant. The projected geodesics of this "minimal" optimal surface coincide with Fermat paths in the 2D media, which is the intrinsic feature of the eikonal equation. This result suggests an idea to treat the growth process itself as a propagation of the wavefronts in the media with certain metrics.

We have derived the metrics of the growing plant's surface from microscopic rules of cells division and have shown that the solution of the eikonal equation describes buckling of tissues of different rigidities. Our results, being purely geometric, rhyme well with a number of energetic approaches
to buckling of thin membranes, where the stiffness is controlled by the effective bending rigidity. We show that presence of a finite scale on which the tissue remains flat, results in negatively curved growing surfaces and the eikonal equation implies absence of real solution at late stages of the growth. Though, an analytical continuation can be constructed and erratic self-similar patterns along the circumference can be obtained. In reality high energetic costs for the profound cell division after bifurcation point would prohibit infinite growth and intense buckling.

Recall, that the right-hand side of the eikonal equation mimics the squared refraction index, \eq{eq:n}, if buckling is interpreted as wavefront propagation in geometric optics. At length of our work it was pointed out, that for the differential growth problem, negative square of refraction index leads to complex solution for $f$. Does complex solution have any physical meaning for growth? We can provide the following speculation. The complex solution appears for the late stages of growth when the finite bending scale of the tissue prohibits formation of very low-wavelength buckling modes. Since in this regime the tissue would experience in-plane deformations, one may improve the geometric model by letting branches to accumulate the "potential energy". Thereby, the analogy between optics and differential growth can be advanced by noting that the negative squared refraction index means absorbtion properties of the media. The propagating wavefront of a moving particle, dissipates the energy in areas where the refraction index is complex-valued. In the differential growth the proliferation of buckling modes may be limited by the energy losses at branches, that would suppress buckling.

The challenging question concerns the possibility to extend our approach to the growth of
three-dimensional objects, for example, of a ball that size $R$ grows faster than $R^2$. In this
case, the redundant material can provoke the surface instabilities. We conjecture that some analogy between the boundary growth and optic wavefronts survives in this case as well.

\section{Rare-event statistics and hyperbolic geometry}
\label{sec3}

The spectral analysis of ensemble of random adjacency matrices at the percolation threshold is
cumbersome since the spectra of the components depend on their topology. For random Erd\H{o}s-R\'enyi graphs, however, almost all components are trees (see e.g. Ref. [\refcite{Janson,book}]). Furthermore, at the percolation point (and certainly below it) the majority of subgraphs are linear chains. It has been shown in Ref. [\refcite{krapivsky}] that in the vicinity of the percolation point linear chains constitute the overwhelming majority of clusters---about 95\%. This has led us to study the spectral density of ensemble of linear chains only. The advantage of this consideration is the possibility of exact treatment which allowed us to uncover the number-theoretic structure of the corresponding density of states. The spectrum of each linear chain is easy to compute, so to get the spectral density, $\rho_{\rm lin}(\lambda)$, of an ensemble of chain-like subgraphs we should know the distribution, $Q_n$, of linear chains of length $n$ at the percolation point. In Ref. [\refcite{krapivsky}] an explicit expression for $Q_n$ has been derived within a kinetic theory approach which allows one to determine the size distribution of generic clusters (irrespective of their topology) and then adopt this framework to the computation of the size distribution of linear chains. As concerns exact results, we have shown in Ref. [\refcite{krapivsky}] that the spectral density, $\rho_{\rm lin}(\lambda)$, in the ensemble of linear chains with the exponential distribution of their lengths demonstrates a very peculiar ultrametric structure related to the so-called "popcorn function" [\refcite{popcorn}].

The "popcorn" (or Thomae) function, $g(x)$, has also many other names: the raindrop function, the countable cloud function, the modified Dirichlet function, the ruler function, etc. It is one of the simplest number-theoretic functions possessing nontrivial fractal structure (another famous example is the everywhere continuous but never differentiable Weierstrass function). The popcorn function is defined on the open interval $x \in (0, 1)$ according to the following rule:
\be
g(x) = \begin{cases} \frac{1}{q} & \mbox{if $x=\frac{p}{q}$, and $(p,q)$ are coprime} \medskip \\
0 & \mbox{if $x$ is irrational} \end{cases}
\ee
The popcorn function $g$ is discontinuous at every rational point because irrationals come
infinitely close to any rational number, while $g$ vanishes at all irrationals. At the same time,
$g$ is continuous at irrationals.

\begin{figure}[ht]
\centerline{\includegraphics[width=10cm]{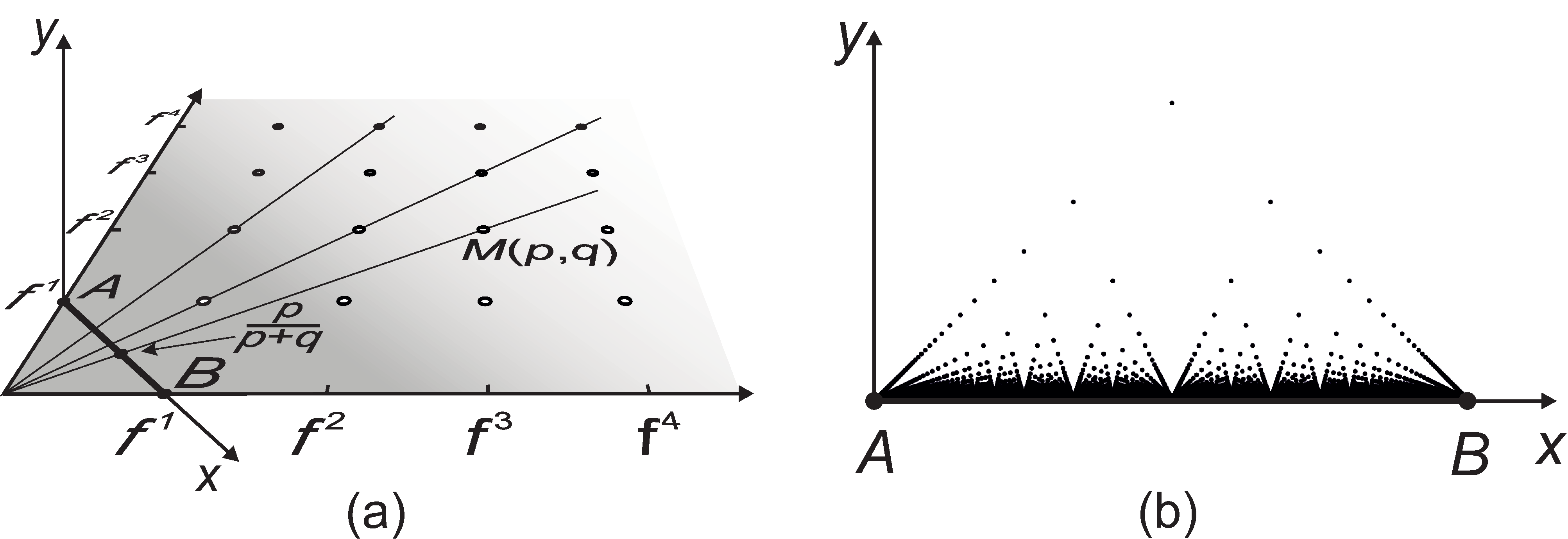}}
\caption{(a) Construction of the Euclid Orchard; (b) Popcorn (Thomae) function.}
\label{fig14}
\end{figure}

One of the most beautiful incarnations of the popcorn function arises in a so-called "Euclid
orchard" representation. Consider an orchard of trees of \emph{unit hights} located at every point $(an,am)$ of the two-dimensional square lattice, where $n$ and $m$ are nonnegative integers
defining the lattice, and $a$ is the lattice spacing, $a = 1/\sqrt{2}$. Suppose we stay on the line $n = 1 - m$ between the points $A(0, a)$ and $B(a, 0)$, and observe the orchard grown in the first
quadrant along the rays emitted from the origin $(0,0)$ -- see the \fig{fig14}.

Along these rays we see only the first open tree with coprime coordinates, $M(ap,aq)$, while all
other trees are shadowed. Introduce the auxiliary coordinate basis $(x,y)$ with the axis $x$ along the segment $AB$ and $y$ normal to the orchard's plane (as shown in the \fig{fig14}a). We set the
origin of the $x$ axis at the point $A$, then the point $B$ has the coordinate $x = 1$. It is a
nice school geometric problem to establish that: (i) having the focus located at the origin, the
tree at the point $M(ap,aq)$ is spotted at the place $x = \frac{p}{p+q}$, (ii) the \emph{visible}
height of this tree is $\frac{1}{p+q}$. In other words, the "visibility diagram" of such a lattice orchard is exactly the popcorn function.

The popcorn correspondence $\frac{p}{q} \to \frac{1}{q}$ arises in the Euclid orchard problem as a
purely geometrical result. However, the same function has appeared as a probability distribution in a plethora of biophysical and fundamental problems, such as the distribution of quotients of reads
in DNA sequencing experiment [\refcite{dna}], quantum $1/f$ noise and Frenel-Landau shift [\refcite{planat}], interactions of non-relativistic ideal anyons with rational statistics parameter in the magnetic gauge approach [\refcite{lundholm}], or frequency of specific subgraphs counting in the protein-protein network of a \textit{Drosophilla} [\refcite{drosophilla}]. Though the extent of similarity with the original popcorn function could vary, and experimental profiles may drastically depend on peculiarities of each particular physical system, a general probabilistic scheme resulting in the popcorn-type manifestation of number-theoretic behavior in nature, definitely survives.

Suppose two random integers, $\phi$ and $\psi$, are taken independently from a discrete probability distribution, $Q_n = f^n$, where $f = 1 - \eps > 0$ is a "damping factor". If $\mathrm{gcd}(p,q) =
1$, then the combination $\nu  = \frac{\phi}{\phi+\psi}$ has the popcorn-like distribution $P(\nu)$ in the asymptotic limit $\eps \ll 1$:
\be
P\left(\nu = \frac{p}{p+q}\right) = \sum_{n=1}^{\infty} f^{n(p+q)} = \frac{(1-\eps)^{p+q}}{1 - (1-\eps)^{p+q}} \approx \frac{1}{\eps(p+q)}
\label{ind}
\ee

The formal scheme above can be understood on the basis of the Euclid orchard construction, if
one would consider a $1+1$ walker on the lattice (see \fig{fig14}a), who performs $\phi$
steps along one axis of the lattice, following by $\psi$ steps along another axis. At every step the walker dies with probability $\eps = 1 - f$. Then, having a number of the walkers starting from the origin of the lattice, one would get an "orchard of walkers", i.e. at every spot $\nu$ on the $x$ axis a fraction of survived walkers $P(\nu)$ would be described exactly by the popcorn function.

In order to have a relevant physical picture, consider a toy model of diblock-copolymer
polymerization. Without sticking to any specific polymerization mechanism, consider an ensemble of diblock-copolymers $AB$, polymerized independently from both ends in a cloud of monomers of
relevant kind (we assume, only $A-A$ and $B-B$ links to be formed). Termination of polymerization
is provided by specific "radicals" of very small concentration, $\eps$: when a radical is attached to the end (irrespectively, $A$ or $B$), it terminates the polymerization at this extremity
forever. Given the environment of infinite capacity, one assigns the probability $f = 1-\eps$ to a monomer attachment at every elementary act of the polymerization. If $N_A$ and $N_B$ are molecular
weights of the blocks $A$ and $B$, then the composition probability distribution in our ensemble,
$P\left(\varphi = \frac{N_A}{N_A+N_B}\right)$, in the limit of small $\eps \ll 1$ is "ultrametric" (see Ref. [\refcite{avetisov}] for the definition of the ultrametricity) and is given by the popcorn function:
\be
P\left(\varphi = \frac{p}{p + q}\right) \approx \frac{1}{\eps(p+q)} \stackrel{def} = \frac{1}{\eps}g(\varphi)
\ee

In the described process we have assumed identical independent probabilities for the monomers of
sorts ("colors") $A$ and $B$ to be attached at both chain ends. Since no preference is implied, one may look at this process as at a homopolymer ("colorless") growth, taking place at two extremities.
For this process we are interested in statistical characteristics of the resulting ensemble of the homopolymer chains. What would play the role of "composition" in this case, or in other words, how
should one understand the fraction of monomers attached at one end? As we show below, the answer is rather intriguing: the respective analogue of the probability distribution is the spectral density
of the ensemble of linear chains with the probability $Q_L$ for the molecular mass distribution,
where $L$ is the length of a chain in the ensemble.

To our point of view, the popcorn function has not yet received decent attention among researchers, though its emergence in various physical problems seems impressive, as we demonstrate below.
Apparently, the main difficulty deals with the discontinuity of $g(x)$ at every rational point,
which often results in a problematic theoretical treatment and interpretation of results for the
underlying physical system. Thus, a natural, physically justified "continuous approximation" to the popcorn function is very demanded.

Below we provide such an approximation, showing the generality of the "popcorn-like" distributions for a class of one-dimensional disordered systems. We demonstrate that the popcorn function can be
constructed on the basis of the modular Dedekind function, $\eta(x+iy)$, when the imaginary part,
$y$, of the modular parameter $z=x+iy$ tends to 0.

\subsection{Spectral density and the popcorn function}
\label{sec31}

The former exercises are deeply related to the spectral statistics of ensembles of linear polymers. In a practical setting, consider an ensemble of noninteracting linear chains with exponential
distribution in their lengths. We claim the emergence of the fractal popcorn-like structure in the spectral density of corresponding adjacency matrices describing the connectivity of elementary
units (monomers) in linear chains.

The ensemble of exponentially weighted homogeneous chains, is described by the bi-diagonal
symmetric $N\times N$ adjacent matrix $B=\{b_{ij}\}$:
\be
B = \left(\begin{array}{ccccc}
0 & x_1 & 0 & 0 & \cdots \\  x_1 & 0 & x_2 & 0 &  \\  0 & x_2 & 0 & x_3 &  \\
0 & 0 & x_3 & 0 &  \\ \vdots &  &  &  & \ddots
\end{array} \right)
\label{eeq:06}
\ee
where the distribution of each $b_{i,i+1}=b_{i+1,i}=x_i$ ($i=1,...,N$) is Bernoullian:
\be
x_i=\left\{\begin{array}{ll} 1 & \mbox{with probability $f$} \medskip \\
0 & \mbox{with probability $\eps = 1-f$} \end{array} \right.
\label{eeq:06a}
\ee
We are interested in the spectral density, $\rho_{\eps}(\lambda)$, of the ensemble of matrices $B$ in the
limit $N\to\infty$. Note that at any $x_k=0$, the matrix $B$ splits into independent blocks. Every
$n\times n$ block is a symmetric $n\times n$ bi-diagonal matrix $A_n$ with all $x_k=1$,
$k=1,...,n$, which corresponds to a chain of length $n$. The spectrum of the matrix $A_n$ is
\be
\lambda_{k,n} = 2\cos\frac{\pi k}{n+1}; \qquad (k=1,...,n)
\label{eeq:06b}
\ee

All the eigenvalues $\lambda_{k,n}$ for $k=1,...,n-1$ appear with the probability $Q_n = f^n$ in
the spectrum of the matrix \eq{eeq:06}. In the asymptotic limit $\eps \ll 1$, one may deduce an
equivalence between the composition distribution in the polymerization problem, discussed in the
previous section, and the spectral density of the linear chain ensemble. Namely, the probability of a composition $\varphi = \frac{p}{p+q}$ in the ensemble of the diblock-copolymers can be precisely mapped onto the peak intensity (the degeneracy) of the eigenvalue $\lambda=\lambda_{p,p+q-1} =
2\cos\frac{\pi p}{p+q}$ in the spectrum of the matrix $B$. In other words, the integer number $k$
in the mode $\lambda_{k,n}$ matches the number of $A$-monomers, $N_A = kz$, while the number of
$B$-monomers matches $N_B = (n + 1 - k)z$, where $z \in N$, in the respective diblock-copolymer.

\begin{figure}[ht]
\centerline{\includegraphics[width=6cm]{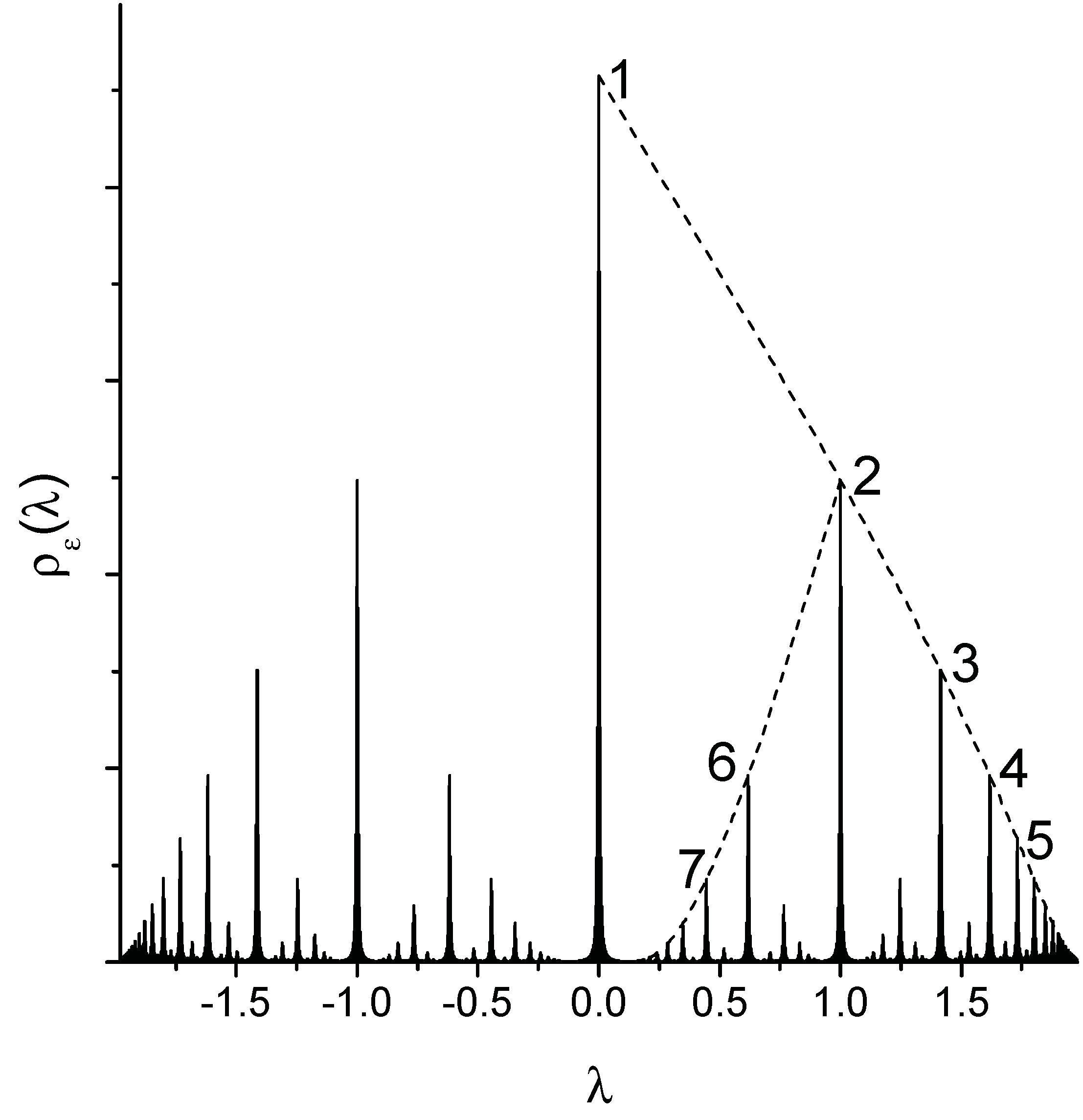}}
\caption{The spectral density $\rho_{\eps}(\lambda)$ for the ensemble of bi-diagonal matrices of
size $N=10^3$ at $f=0.7$. The regularization parameter $\eps$ is taken $\eps=2\times 10^{-3}$.}
\label{fig15}
\end{figure}

To derive $\rho_{\eps}(\lambda)$ for arbitrary values of $\eps$, let us write down the spectral density of the ensemble of $N\times N$ random matrices $B$ with the bimodal distribution of the elements as a resolvent:
\begin{multline}
\rho_\eps(\lambda) = \lim_{N\to\infty}\Big< \sum_{k=1}^{n} \delta(\lambda-\lambda_{kn}) \Big>_{Q_n} = \lim_{N\to\infty \atop y\to+0} y\; \im\, \Big< G_n(\lambda - iy) \Big>_{Q_n} \\ =
\lim_{N\to\infty \atop y\to+0} y \sum_{n=1}^N Q_n\; \im\, G_n(\lambda - iy)
\label{rh}
\end{multline}
where $\la ...\ra_{Q_n}$ means averaging over the distribution $Q_n=(1-\eps)^n$, and the following
regularization of the Kronecker $\delta$-function is used:
\be
\delta(\xi) = \lim_{y\to+0} \im \frac{y}{\xi- iy}
\label{delt}
\ee
The function $G_n$ is associated with each particular gapless matrix $B$ of $n$ sequential "1" on
the sub-diagonals,
\be
G_n(\lambda-iy) = \sum_{k=1}^n \frac{1}{\lambda-\lambda_{k,n}-iy}
\label{green}
\ee
Collecting \eq{eeq:06b}, \eq{rh} and \eq{green}, we find an explicit expression for the density of
eigenvalues:
\be
\rho_\eps(\lambda) = \lim_{N\to\infty \atop y\to+0} y \sum_{n=1}^{N} (1-\eps)^n
\sum_{k=1}^n\frac{y}{\left(\lambda-2\cos\frac{\pi k}{n+1}\right)^2+y^2}
\label{eeq:10}
\ee
The behavior of the inner sum in the spectral density in the asymptotic limit $y \to 0$ is easy to
understand: it is $\frac{1}{y}$ at $\lambda=2\cos{\frac{\pi k}{n+1}}$ and zero otherwise. Thus, one can already infer a qualitative similarity with the popcorn function. It turns out, that the
correspondence is quantitative for $\eps = 1 - f \ll 1$. Driven by the purpose to show it, we
calculate the values of $\rho_\eps(\lambda)$ at the peaks, i.e. at rational points $\lambda =
2\cos{\frac{\pi p}{p+q}}$ with $\mathrm{gcd}(p, q) = 1$ and end up with the similar geometrical
progression, as for the case of diblock-copolymers problem \eq{ind}:
\begin{multline}
\rho_\eps \left(\lambda = 2\cos{\frac{\pi p}{p+q}}\right) = \sum_{s=1}^{\infty} (1-\eps)^{(p+q)s-1}
\\ = \frac{(1-\eps)^{p+q-1}}{1 - (1-\eps)^{p+q}}\Bigg|_{\eps\to 0} \approx \frac{1}{\eps(p+q)}
\stackrel{def} = g\left(\frac{1}{\pi}\arccos{\frac{\lambda}{2}}\right)
\label{eeq:115}
\end{multline}
The typical sample plot $\rho_{\eps}(\lambda)$ for $f=0.7$ computed numerically via \eq{eeq:10}
with $\eps=2\times 10^{-3}$ is shown in the \fig{fig15} for $N=10^3$.

\subsection{Enveloping curves and tails of the eigenvalues density}
\label{sec32}

\begin{figure}[ht]
\centerline{\includegraphics[width=10cm]{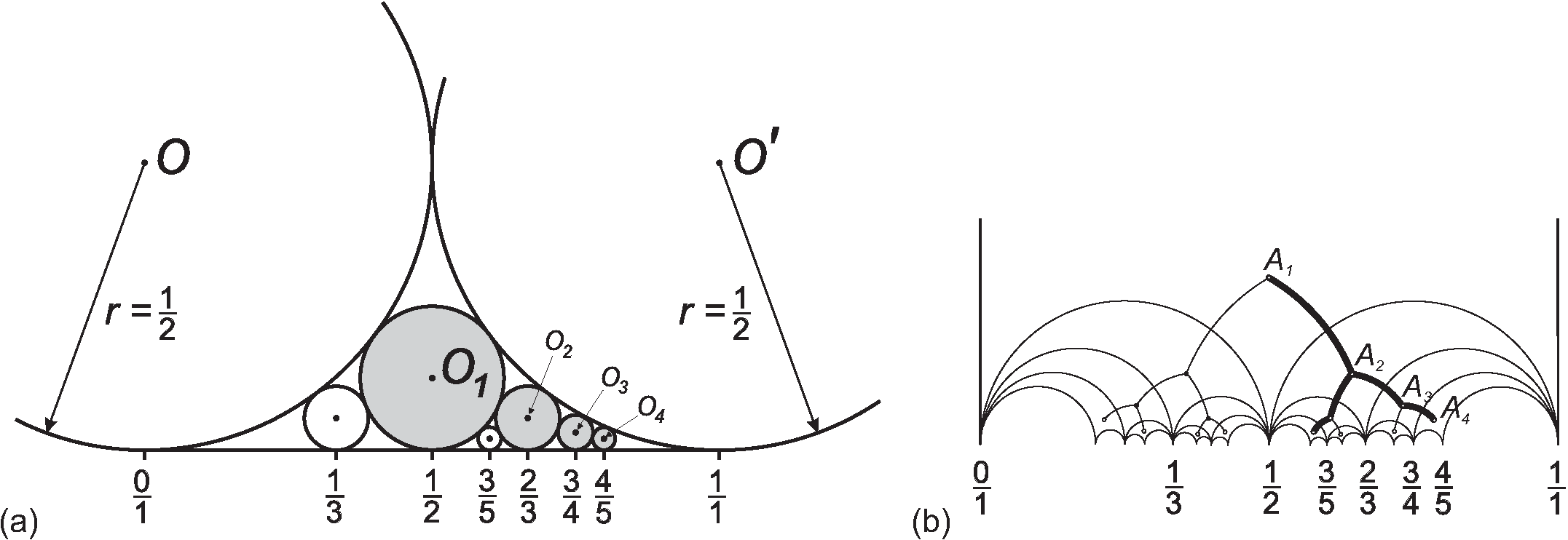}}
\caption{Ford circles as illustration of the Farey sequence construction: (a) Each circle touches
two neighbors (right and left) and the segment. The position of newly generated circle is
determined via the $\oplus$ addition: $\frac{p_{k-1}}{q_{k-1}} \oplus \frac{p_{k+1}}{q_{k+1}}=
\frac{p_{k-1}+p_{k+1}}{q_{k-1}+q_{k+1}}$; (b) The same Farey sequence generated by sequential
fractional-linear transformations of the fundamental domain of the modular group $SL(2,\mathbb{Z})$. }
\label{fig16}
\end{figure}

Below we pay attention to some number-theoretic properties of the spectral density of the argument $-\lambda$, since in this case the correspondence with the composition ratio is precise. One can
compute the enveloping curves for any monotonic sequence of peaks depicted in \fig{fig15}, where
we show two series of sequential peaks: $S_1=\{$1--2--3--4--5--...$\}$ and $S_2=\{$2--6--7-- ...$\}$. Any monotonic sequence of peaks corresponds to the set of eigenvalues $\lambda_{k,n}$ constructed on the basis of a Farey sequence [\refcite{farey}]. For example, as shown below, the peaks in the series $S_1$ are located at:
$$
\lambda_k = -\lambda_{k,k} =-2\cos\frac{\pi k}{k+1}, \qquad (k=1,2,...)
$$
while the peaks in the series $S_2$ are located at:
$$
\lambda_{k'} = -\lambda_{k',2k'-2} = -2\cos \frac{\pi k'}{2k'-1}, \qquad (k'=2,3,...)
$$
Positions of peaks obey the following rule: let $\{\lambda_{k-1},\, \lambda_k,\, \lambda_{k+1}\}$
be three consecutive monotonically ordered peaks (e.g., peaks 2--3--4 in \fig{fig15}), and let
$$
\lambda_{k-1}=-2\cos \frac{\pi p_{k-1}}{q_{k-1}}, \quad \lambda_{k+1}=-2\cos \frac{\pi
p_{k+1}}{q_{k+1}}
$$
where $p_k$ and $q_k$ ($k=1,...,N$) are coprimes. The position of the intermediate peak,
$\lambda_k$, is defined as
\be
\lambda_{k}=-2\cos \frac{\pi p_{k}}{q_{k}}; \qquad \frac{p_{k}}{q_{k}} = \frac{p_{k-1}}{q_{k-1}}
\oplus \frac{p_{k+1}}{q_{k+1}} \equiv \frac{p_{k-1}+p_{k+1}}{q_{k-1}+q_{k+1}}
\label{eeq:11}
\ee
The sequences of coprime fractions constructed via the $\oplus$ addition are known as Farey
sequences. A simple geometric model behind the Farey sequence, known as Ford circles [\refcite{ford, coxeter}], is depicted in \fig{fig16}a. In brief, the construction goes as follows. Take the segment $[0,1]$ and draw two circles $O_1$ and $O_2$ both of radius $r=\frac{1}{2}$, which touch each other, and the segment at the points 0 and 1. Now inscribe a new circle $O_3$ touching $O_1$, $O_2$ and $[0,1]$. Where is the position of the new circle along the segment? The generic recursive algorithm constitutes the Farey sequence construction. Note that the same Farey sequence can be sequentially generated by fractional-linear transformations (reflections with respect to the arcs) of the fundamental domain of the modular group $SL(2,\mathbb{Z})$ -- the triangle lying in the upper halfplane $\im z>0$ of the complex plane $z$ (see \fig{fig16}b).

Consider the main peaks series, $S_1=\{$1--2--3--4--5--...$\}$. The explicit expression for their
positions reads as:
\be
\lambda_k = -2\cos\frac{\pi k}{k+1}; \qquad k = 1, 2, ...
\ee
One can straightforwardly investigate the asymptotic behavior of the popcorn function in the limit
$k \to \infty$. From \eq{eeq:115} one has for arbitrary $f < 1$ the set of parametric equations:
\be
\left\{\begin{array}{l}
\disp \rho_\eps(\lambda_k) = \frac{f^{k}}{1 - f^{k+1}}\Big|_{k\gg 1} \approx f^{k} \\
\disp \lambda_k=-2\cos\frac{\pi k}{k+1}\Big|_{k\gg 1} \approx 2 - \frac{\pi^2}{k^2}
\end{array}\right.
\label{asm}
\ee
From the second equation of \eq{asm}, we get $k\approx \frac{\pi}{\sqrt{2-\lambda}}$. Substituting
this expression into the first one of \eq{asm}, we end up with the following asymptotic behavior of
the spectral density near the spectral edge $\lambda\to 2^-$:
\be
\rho_{\eps}(\lambda) \approx \exp\left(\frac{\pi \ln f}{\sqrt{2-\lambda}}\right) \qquad (0<f<1)
\label{asm2}
\ee
The behavior \eq{asm} is the signature of the Lifshitz tail typical for the 1D Anderson
localization:
\be
\rho_{\eps}(E)\approx e^{-C E^{-D/2}};
\ee
where $E= 2-\lambda$ and $D=1$.

\subsection{From popkorn to Dedekind $\eta$-function}

The popcorn function has discontinuous maxima at rational points and continuous valleys at irrationals. We show in this section, that the popcorn function can be regularized on the basis of
the everywhere continuous Dedekind function $\eta(x+iy)$ in the asymptotic limit $y\to 0$.

The famous Dedekind $\eta$-function is defined as follows (this function has appeared already in \eq{eq:ded1}):
\be
\eta(z)=e^{\pi i z/12}\prod_{n=0}^{\infty}(1-e^{2\pi i n z})
\label{eeq:dedeta}
\ee
The argument $z=x+iy$ is called the modular parameter and $\eta(z)$ is defined for $\im z>0$ only.
The Dedekind $\eta$-function is invariant with respect to the action of the modular group
$SL(2,\mathbb{Z})$:
\be
\begin{array}{l}
\eta (z+1)=e^{\pi i z/12}\;\eta(z) \medskip \\ \eta\left(-\frac{1}{z}\right) = \sqrt{-i}\; \eta(z)
\end{array}
\label{2}
\ee
And, in general,
\be
\eta\left(\frac{az+b}{cz+d}\right) = \omega(a,b,c,d)\; \sqrt{c z + d}\; \eta(z)
\label{3}
\ee
where $ad-bc=1$ and $\omega(a,b,c,d)$ is some root of 24th degree of unity [\refcite{dedekind}].

It is convenient to introduce the following "normalized" function
\be
h(z) = |\eta(z)| (\im z)^{1/4}
\label{eeq:15}
\ee
The real analytic Eisenstein series $E(z, s)$ is defined in the upper half-plane, $H = \{z: \im(z)
> 0\}$ for $\re(s) > 1$ as follows:
\be
E(z,s) = \frac{1}{2}\sum_{\{m,n\} \in \mathbb{Z}^2 \backslash \{0, 0\}} \frac{y^s}{|mz + n|^{2s}};
\qquad z = x+iy
\ee
This function can be analytically continued to all $s$-plane with one simple pole at $s=1$. Notably
it shares the same invariance properties on $z$ as the Dedekind $\eta$-function. Moreover, $E(s,
z)$, as function of $z$, is the $SL(2,\mathbb{Z})$--automorphic solution of the hyperbolic Laplace equation:
$$
-y^2 \left(\frac{\partial^2}{\partial x^2}+\frac{\partial^2}{\partial y^2}\right)
E(z, s) = s(1-s)\; E(z, s)
$$

The Eisenstein series is closely related to the Epstein $\zeta$-function, $\zeta(s, Q)$, namely:
\be
\zeta(s, Q) = \sum_{\{m,n\} \in \mathbb{Z}^2 \backslash \{0, 0\}} \frac{1}{Q(m, n)^s} =
\frac{2}{d^{s/2}}E(z, s),
\label{epstein}
\ee
where $Q(m, n) = am^2 + 2bmn + cn^2$ is a positive definite quadratic form, $d=ac-b^2 > 0$, and
$\disp z = \frac{-b+i\sqrt d}{a}$. Eventually, the logarithm of the Dedekind $\eta$-function is
known to enter in the Laurent expansion of the Epstein $\zeta$-function. Its residue at $s=1$ has
been calculated by Dirichlet and is known as the first Kronecker limit formula
[\refcite{epstein,siegel,motohashi}]. Explicitly, it reads at $s\to 1$:
\be
\zeta(s, Q) = \frac{\pi}{\sqrt d} \frac{1}{s-1} + \frac{2\pi}{\sqrt d}\left(\gamma +
\ln\sqrt{\frac{a}{4d}} - 2\ln|\eta(z)|\right) + O(s-1)
\label{eeq:zeta}
\ee
Equation \eq{eeq:zeta} establishes the important connection between the Dedekind $\eta$-function and the respective series, that we substantially exploit below.

\begin{figure}[ht]
\centerline{\includegraphics[width=9cm]{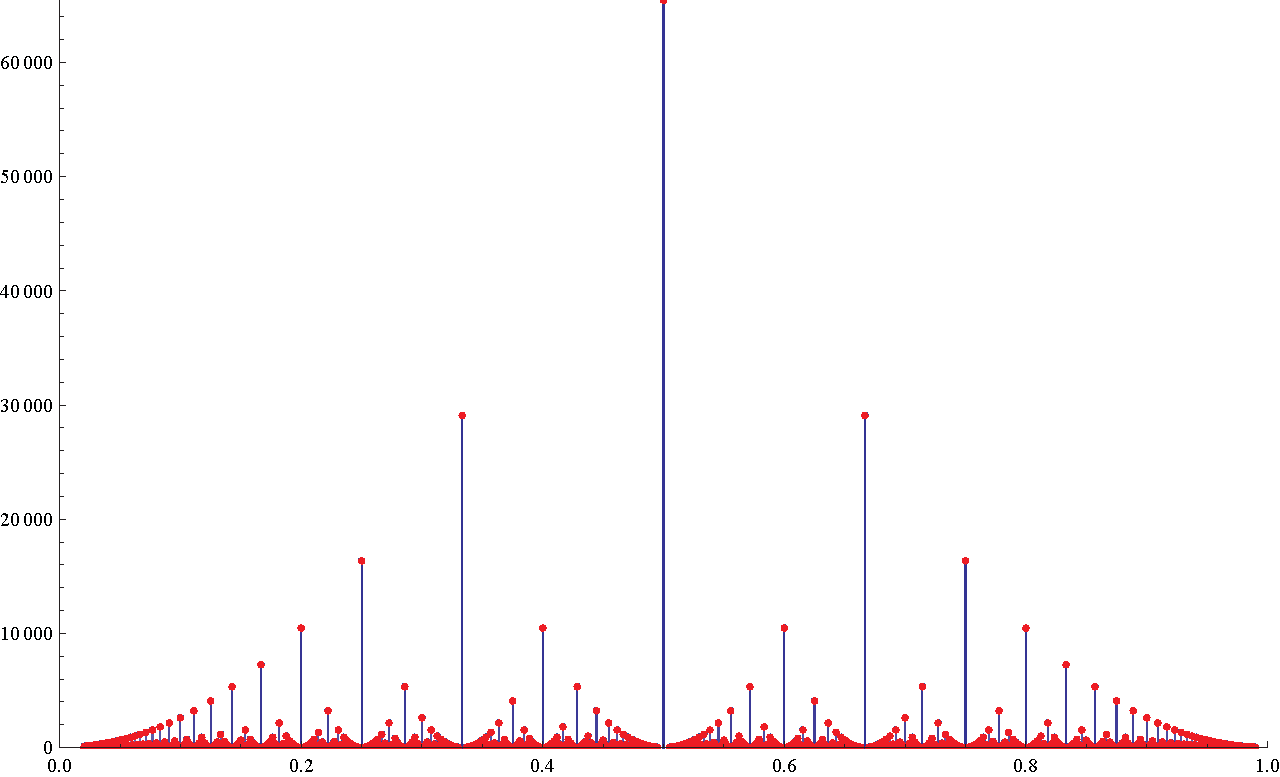}}
\caption{Plots of everywhere continuous $f_1(x) = -\ln|\eta(x+i\eps)|$ (blue) and discrete
$f_2(x) = \frac{\pi}{12\eps} g^2(x)$ (red) for $\eps = 10^{-6}$ at rational points in $0<x<1$.}
\label{fig17}
\end{figure}

Consider an arbitrary quadratic form $Q'(m,n)$ with unit determinant. Since $d=1$, it can be
written in new parameters $\{a, b, c\} \to \{x = \frac{b}{c}, \eps = \frac{1}{c}\}$ as follows:
\be
Q'(m, n) = \frac{1}{\eps}(x m - n)^2 + \eps m^2
\label{q}
\ee
Applying the first Kronecker limit formula to the Epstein function with \eq{q} and $s = 1 + \tau$,
where $\tau \ll 1$, but finite one gets:
\be
\zeta(s,Q') = \frac{\pi}{s-1} + 2\pi \left(\gamma + \ln\sqrt{\frac{1}{4\eps}} - 2\ln|\eta(x+i
\eps)|\right) + O(s-1)
\label{zeta1}
\ee
On the other hand, one can make use of the $\eps$-continuation of the Kronecker $\delta$-function,
\eq{delt}, and assess $\zeta(1 + \tau, Q')$ for small $\tau \ll 1$ as follows:
\begin{multline}
\zeta(1 + \tau, Q') \approx \frac{1}{\eps}\sum_{\{m,n\} \in \mathbb{Z}^2 \backslash \{0, 0\}}
\frac{\eps^2}{\left(x m - n\right)^{2} + \eps^2m^2} \\ = \frac{2}{\eps}\lim_{N\to\infty}
\sum_{m=1}^{N} \sum_{n=1}^{N} \frac{1}{m^2}\delta\left(x - \frac{n}{m}\right) \equiv \theta(x)
\end{multline}
where $x \in (0, 1)$ and the factor 2 reflects the presence of two quadrants on the $\mathbb{Z}^2$-lattice that contribute jointly to the sum at every rational points, while $\theta$
assigns 0 to all irrationals. At rational points $\theta\left(\frac{p}{q}\right)$ can be calculated
straightforwardly:
\be
\theta\left(\frac{p}{q}\right) = \frac{2}{\eps} \sum_{m|q}^{\infty} \frac{1}{m^2} =
\frac{\pi^2}{3 \eps q^2}
\label{eeq:14}
\ee
Comparing \eq{eeq:14} with the definition of the popcorn function, $g$, one ends up with the
following relation at the peaks:
\be
g\left(\frac{p}{q}\right) = \sqrt{\frac{3\eps}{\pi^2}\theta\left(\frac{p}{q}\right)}
\label{popc}
\ee
Eventually, collecting \eq{zeta1} and \eq{popc}, we may write down the regularization of the
popcorn function by the Dedekind $\eta (x + i\eps)|_{\eps \to 0}$ in the interval $0 < x < 1$:
\be
g(x) \approx \sqrt{-\frac{12\eps}{\pi} \ln|\eta(x+i\eps)| - o\left(\eps\ln\eps\right)} \Bigg|_{\eps
\to 0}
\label{result}
\ee
or
\be
-\ln|\eta(x+i\eps)|_{\eps \to 0} = \frac{\pi}{12\eps} g^2(x) + O(\ln\eps)
\label{result1}
\ee
Note, that the asymptotic behavior of the Dedekind $\eta$-function can be independently derived
through the duality relation, [\refcite{vas,krapivsky}]. However, such approach leaves in the dark the underlying structural equivalence of the popcorn and $\eta$ functions and their series
representation on the lattice $\mathbb{Z}^2$. In the \fig{fig17} we show two discrete plots of
the left and the right-hand sides of \eq{result1}.

Thus, the spectral density of ensemble of linear chains, \eq{eeq:115}, in the regime $\eps \ll 1$
is expressed through the Dedekind $\eta$-function as follows:
\be
\rho_\eps (\lambda) \approx \sqrt{-\frac{12\eps}{\pi} \ln\Bigg|\eta\left(\frac{1}{\pi}\arccos{\frac{\lambda}{2}}+i\eps\right)\Bigg|}
\ee

\section{Ultrametricity in physical systems}
\label{sec33}

\subsection{Modular geometry and phyllotaxis}

We have discussed the number-theoretic properties of distributions appearing in physical systems
when an observable is a quotient of two independent exponentially weighted integers. The spectral
density of ensemble of linear polymer chains distributed with the law $f^L$ ($0<f<1$), where $L$ is the chain length, serves as a particular example. At $f\to 1$, the spectral density can be
expressed through the discontinuous and non-differentiable at all rational points, Thomae
("popcorn") function. We suggest a continuous approximation of the popcorn function, based on the
Dedekind $\eta$-function near the real axis.

Analysis of the spectrum at the edges reveals the Lifshitz tails, typical for the 1D Anderson localization. The non-trivial feature, related to the asymptotic behavior of the shape of the spectral density of the adjacency matrix, is as follows. The main, enveloping, sequence of
peaks $1-2-3-4-5...$ in the \fig{fig15} has the asymptotic behavior $\rho(\lambda)\sim
q^{\pi/\sqrt{2-\lambda}}$ (at $\lambda\to 2^-$) typical for the 1D Anderson localization, however any internal subsequence of peaks, like $2-6-7-...$, has the behavior $\rho'(\lambda)\sim
q^{\pi/|\lambda-\lambda_{cr}|}$ (at $\lambda\to \lambda_{cr}$) which is reminiscent of the Anderson localization in 2D.

We would like to emphasize that the ultrametric structure of the spectral density is ultimately
related to number-theoretic properties of modular functions. We also pay attention to the
connection of the Dedekind $\eta$-function near the real axis to the invariant measures of some
continued fractions studied by Borwein and Borwein in 1993 in Ref. [\refcite{borwein}]. The notion of ultrametricity deals with the concept of hierarchical organization of energy landscapes
[\refcite{mez,avetisov}]. A complex system is assumed to have a large number of metastable states
corresponding to local minima in the potential energy landscape. With respect to the transition
rates, the minima are suggested to be clustered in hierarchically nested basins, i.e. larger basins consist of smaller basins, each of these consists of even smaller ones, \emph{etc}. The basins of
local energy minima are separated by a hierarchically arranged set of barriers: large basins are
separated by high barriers, and smaller basins within each larger one are separated by lower
barriers. Ultrametric geometry fixes taxonomic (i.e. hierarchical) tree-like relationships between elements and, speaking figuratively, is closer to Lobachevsky geometry, rather to the Euclidean
one.

\begin{figure}[ht]
\centerline{\includegraphics[width=10cm]{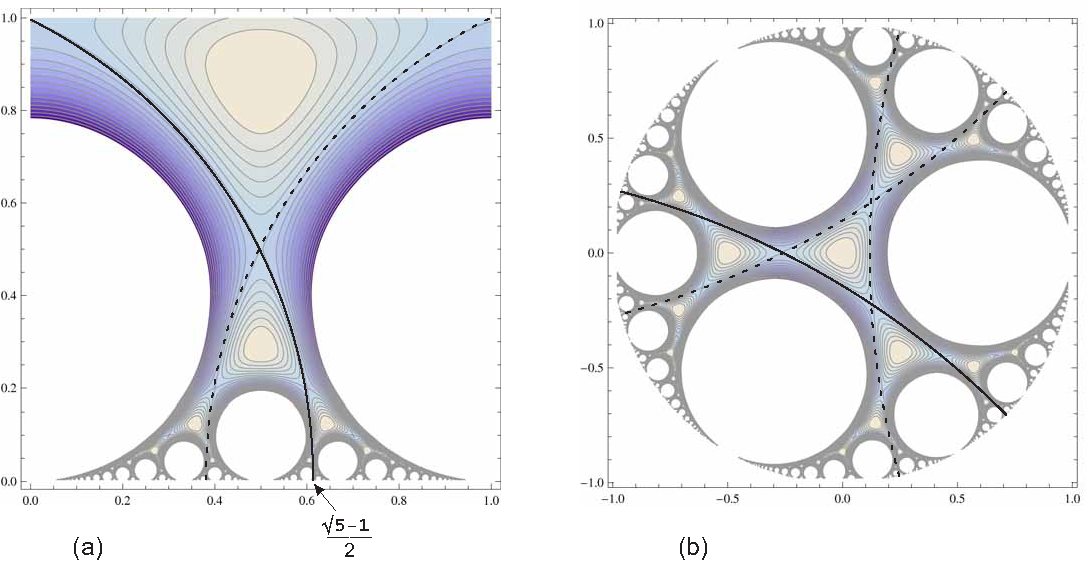}}
\caption{Cut ($f(z)\in[0.985, 1.0]$) of the contour plot of the function $f(z)$ (see
\eq{eeq:15}): a) the representation in the upper half-plane, $z=x+iy$; b) the representation in the unit disc obtained by the conformal mapping of the upper half-plant to the unit disc.}
\label{fig18}
\end{figure}

Let us remind that in the Section \ref{sec2} we argued that buckling of a leaf of some plants (like lettuce or spinach) is related with isometric embedding of hyperbolic graphs (Cayley trees) into the 3D Euclidean space. The buckling is described by a function, $g(z)$, the Jacobian of the conformal transformation. For a regular 3-branching Cayley tree, the function $g(z)$ can be expressed via the Dedekind $\eta$-function [\refcite{voit,vas}]. It has been noted in Ref. [\refcite{vas}] that properly normalized function $g(z)$, namely, $f(z)=C^{-1} |\eta(z)| (\im z)^{1/4}$, defined in Eq.\eq{eeq:15}, can be regarded as a continuous tree isometrically embedded into the half-plane $\im z>0$. The cut of the function $f(z)\in [0.985,1.0]$ is shown in \fig{fig18}a, while in \fig{fig18}b the same function is replotted in polar coordinates, obtained via the conformal mapping of the upper half-plane $\im z>0$ to the unit disc. Note the striking similarity of \fig{fig18}a with the set of touching Ford circles (\fig{fig17}). In fact, the $x$-coordinates of centers of all lacunas in \fig{fig18}a obey the Farey construction.

Another well-known manifestation of number theory in natural science arises in the context of
phyllotaxis [\refcite{phyllo}]. The connection of the cell packing with the Farey sequences has been observed long time ago. It was not clear, however, why nature selects the Fibonacci sequence among other possible Farey sequences. A tantalizing answer to this question has been given by L. Levitov, see Ref. [\refcite{levitov1}]. He proposed an ``energetic" approach to the phyllotaxis, suggesting that the development of a plant is connected with an effective motion along the \emph{geodesics} on the surface associated with the energetic relief of growing plant.

Intriguingly, the energetic relief considered in Ref. [\refcite{levitov1}] coincides with the relief of the function $f(z)$ depicted in \fig{fig18} (to the best of our knowledge, the Dedekind $\eta$ function has newer been discussed in this context). The black arcs drawn in \fig{fig18}a,b are parts of semicircles orthogonal to the boundaries and represent the geodesics of the surface
$f(z)$, which are \emph{open level lines} (by the symmetry, there are only two such level lines in
\fig{fig18}a and three in \fig{fig18}b). The relief of the function $f(z)$ shown in the
\fig{fig18}a emerged previously in course of investigation of metric properties of ultrametric
spaces [\refcite{vas}]. The best rational approximation of the boundary point, $\frac{\sqrt{5}-1}{2}$, at which the solid geodesics in \fig{fig18}a terminates, is given by the continued fraction expansion
\be
\frac{\sqrt{5}-1}{2}=\frac{1}{\disp 1+\frac{1}{\disp 1+\frac{1}{\disp 1+...}}},
\label{eq:cont-fr}
\ee
cut at some level $k$. Cutting \eq{eq:cont-fr} at sequential $k$, we get the set of fractions
constituting the interlacing Fibonacci sequences, $\{F_k\}$, in nominators and denominators:
$$
\frac{1}{1},\; \frac{1}{2},\; \frac{2}{3},\; \frac{3}{5},\;
\frac{5}{8},...,\frac{F_{k-1}}{F_{k}},\frac{F_k}{F_{k+1}}
$$
Thus, the  normalized $\eta$-function, $f(z)$, plays a role of the energy relief of a growing
plant.

The Fibonacci numbers are the benchmarks, which fix the connection between ultrametric and
hyperbolic geometry, since, on one hand they appear as discrete symmetries of the hyperbolic space,
and on the other hand, being a subset of $p$-adic numbers, they uniquely parameterize the distances in the ultrametric space [\refcite{krapivsky}].

\begin{figure}[ht]
\centerline{\includegraphics[width=3.16cm]{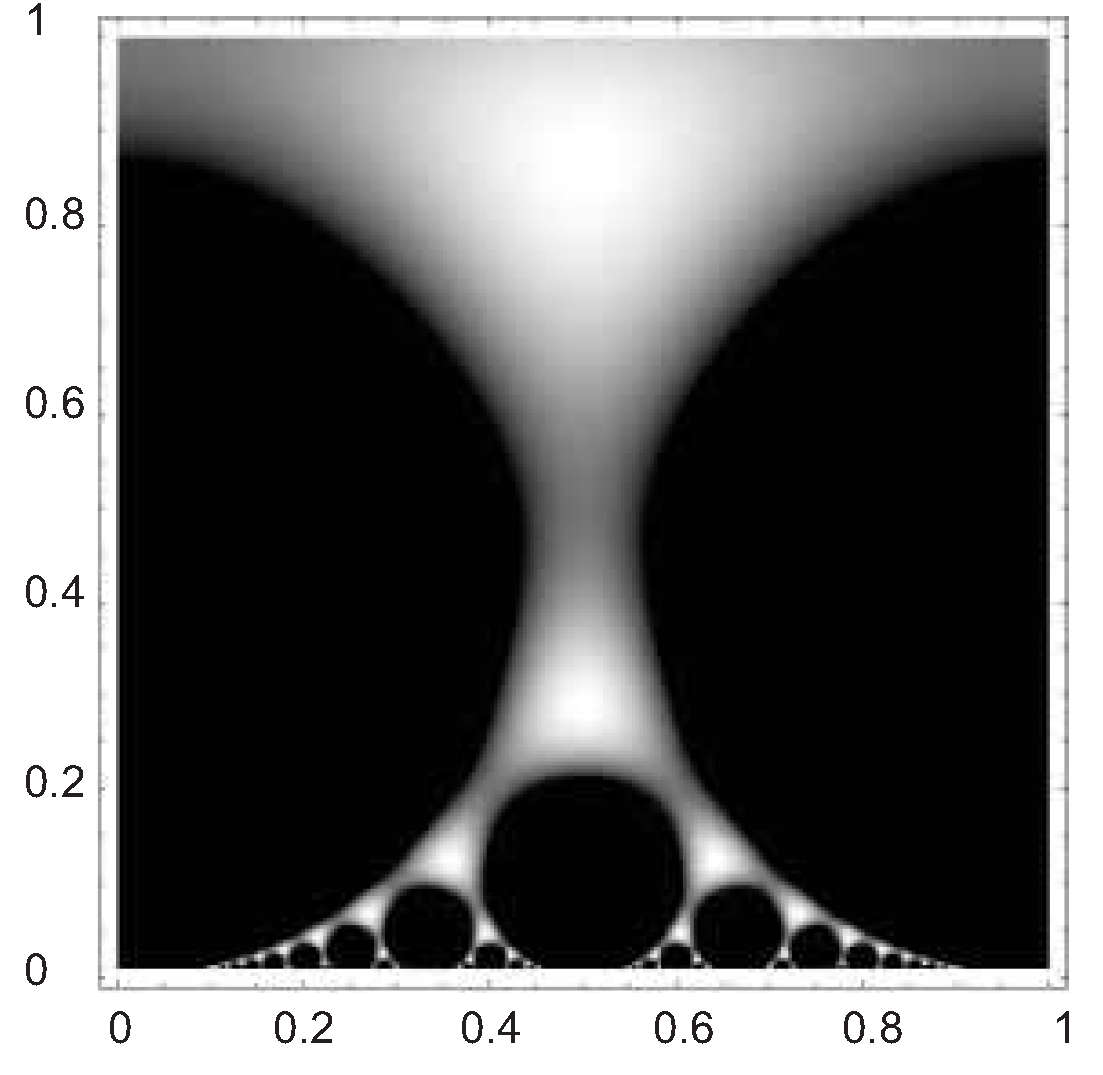}
\includegraphics[width=3.8cm]{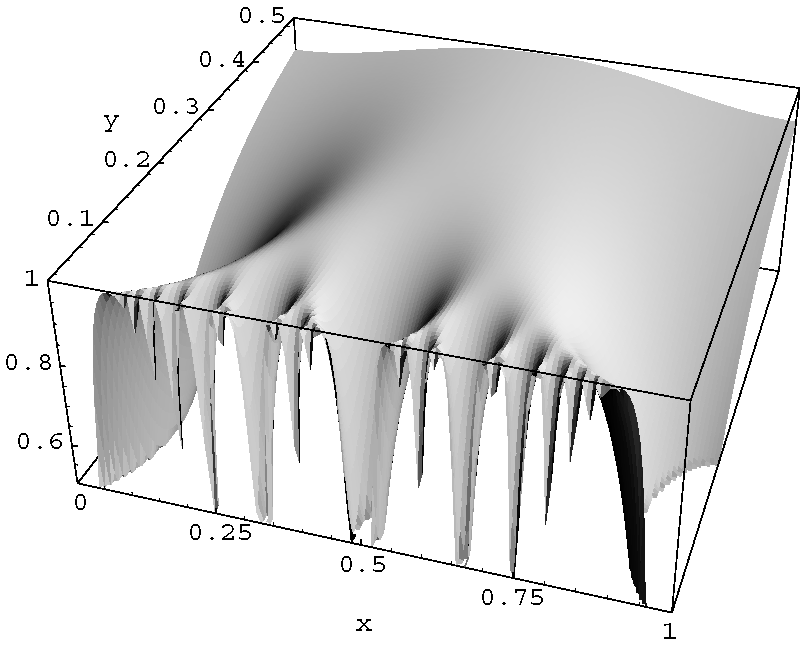}
\includegraphics[width=4cm]{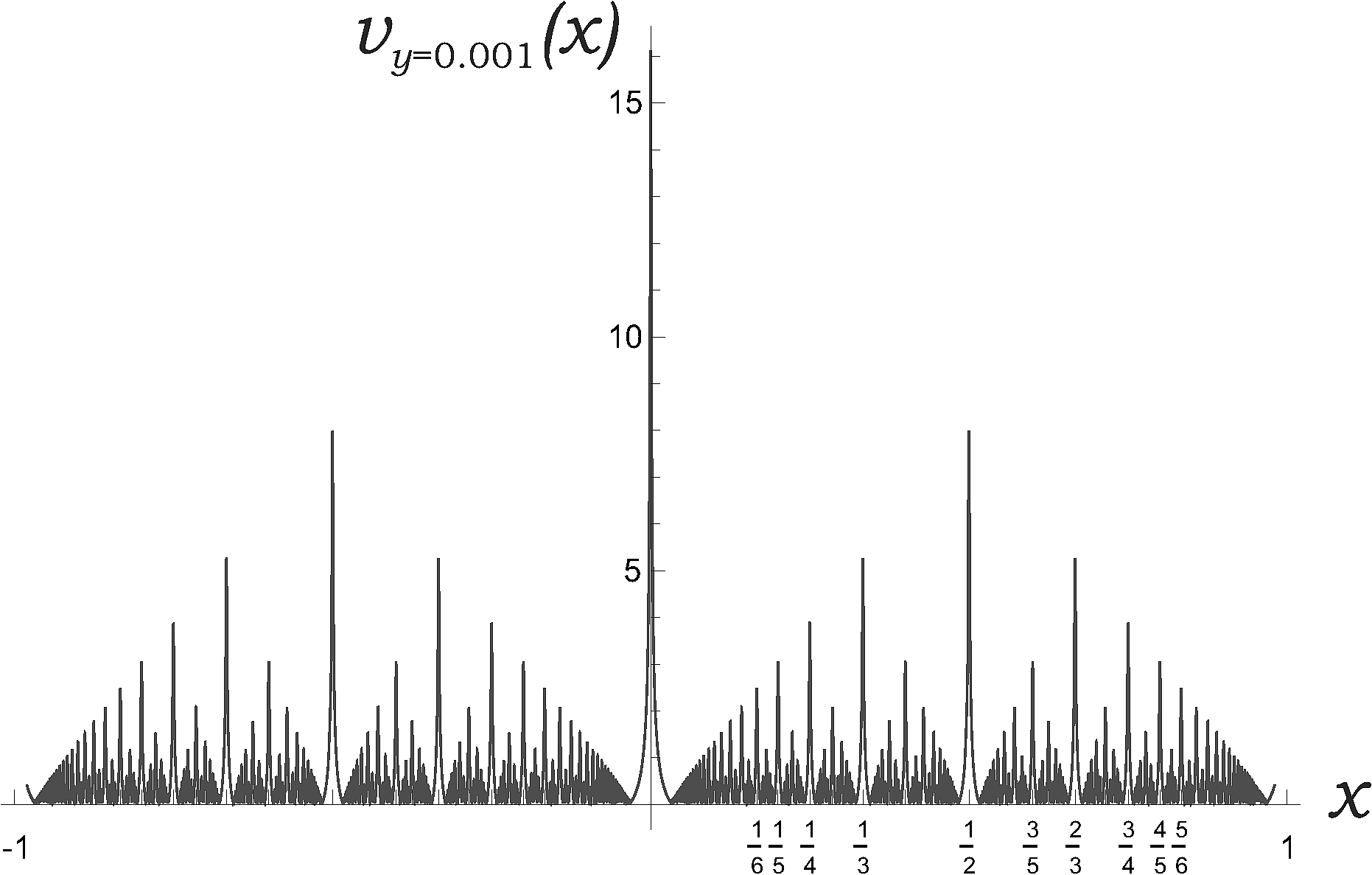}}
\caption{(a) Density plot of the function $f(z)$ in the rectangle
$\{0\le \re z\le 1,\; 0.01\le \im z\le 1\}$; (b) 3D relief of the function $f(z)$ in the rectangle
$\{0\le \re z\le 1,\; 0.04\le \im z\le 0.5\}$; (c) Plot of the function $v_y(x)$ at y=0.001.}
\label{fig19}
\end{figure}

In Ref. [\refcite{vas}] we have constructed the "continuous" analog of the standard 3-branching Cayley tree by means of modular functions and have analyzed the structure of the barriers separating the neighboring valleys. In particular, we have shown that these barriers are ultrametrically organized. The main ingredient of our construction is the function $f(z)$ defined in \eq{eeq:15}. The normalization constant $C$ is chosen to fix the maximal value of $f(z)$ equal to 1: $0<f(z)\le 1$ for any $z$ in the upper half-plane $\im z>0$. The function $f(z)$ has the following property: all the solutions of the equation $f(z)-1=0$ define all the coordinates of the 3-branching Cayley tree isometrically embedded into the upper half-plane ${\cal H}(z|{\rm Im}\,z>0)$ with hyperbolic metric. The density plot of the function $f(z)$ in the region $\{0\le {\rm Re}\,z \le 1,\; 0.04\le \im z\le 1\}$ and its 3D relief are shown in the \fig{fig19}a,b. The \fig{fig19}b clearly displays the "continuous" tree-like structure of hills separated by the valleys. Consider now the function
\be
v_y(x)=\sqrt{-\ln f(x+iy)}; \qquad \im z>0
\label{eq:v}
\ee
The typical shape of $v_{y={\rm const}}(x)$ is shown in the figure \fig{fig19}c for $y=0.001$. The
function $v_y(x)$ at any fixed $y$ obeys the strong triangle inequality condition and hence
possesses the ultrametric organization of barriers separating the valleys. At $y\to 0^+$ the
function $v(x)$ exhibits the "linearity" in sequentially increasing and decreasing barriers similar to the well-known Thomae ("popcorn") function [\refcite{thomae}].

\begin{figure}[ht]
\centerline{\includegraphics[width=11cm]{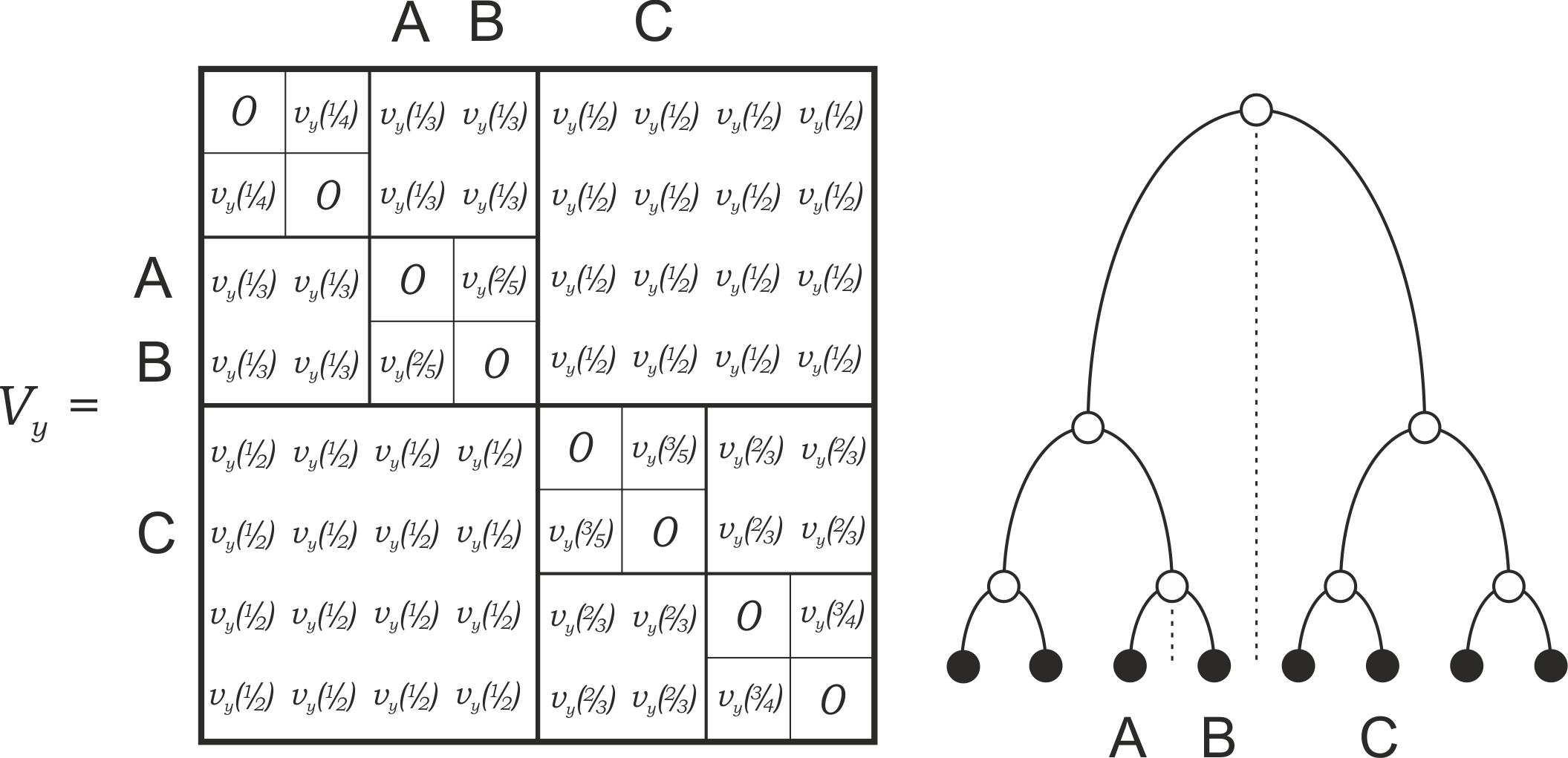}}
\caption{(a) Modular analog of the hierarchical Parisi matrix, (b) The tree-like hierarchical organization of barriers, separating states represented by black dots.}
\label{fig20}
\end{figure}

Using the function $v_y(x)$ we can construct the modular analog of the "Parisi matrix", $V_y$ (the typical object in the spin glass theory [\refcite{mez}]), which for first 3 hierarchical levels can be written in the form shown in figure \fig{fig20}(left). Corresponding tree-like structure with hierarchical (ultrametric) organization of barriers, separating transitions form different states represented by black dots, is depicted in the \fig{fig20}(right).

The function $v_y(x)$, defined on the interval $0<x<1$, has the properties borrowed from the
structure of the underlying modular group $PSL(2,\mathbb{Z})$ acting in the half--plane ${\cal
H}(z|y>0)$ [\refcite{magnus,beardon}]. In particular: (i) the local maxima of $v_y(x)$ are located at the rational points; (ii) the highest barrier on a given interval $\Delta x = [x_1,x_2]$ is located at a rational point $\frac{p}{q}$ with the lowest denominator $q$. On a given interval $\Delta x = [x_1,x_2]$ there is only one such point. Locations of barriers with consecutive heights on the interval $\Delta x$ are organized according to the group operation $\frac{p_1}{q_1}\oplus \frac{p_2}{q_2} = \frac{p_1+p_2}{q_1+q_2}$, constituting the Farey sequence discussed in the main
text of the paper. For example, the height of the barrier separating points $A$ and $B$ is $v_y(2/3)$, while the height of barriers separating points $A$ and $C$, or $B$ and $C$ is $v_y(1/2)$.

The array shown in the \fig{fig20} can be considered as a "continous" analog of the Parisi matrix and could play a role in problems where the ultrametricity should be consistent with metric structure of the target space. 

\subsection{Spectral statistics of ensembles of trees and star-like graphs}

One can gain the information about topological and statistical properties of polymers in solutions
by measuring their relaxation spectra \cite{brouwer2011spectra}. A rough model of an individual
polymer molecule of any topology is a set of monomers (atoms) connected by elastic strings. If
deformations of strings are small, the response of the molecule on external excitation is harmonic
according to the Hooke's law. The relaxation modes are basically determined by the so-called
Laplacian matrix (defined below) of the molecule.

\begin{figure}[ht]
\centerline{\includegraphics[width=5.2cm]{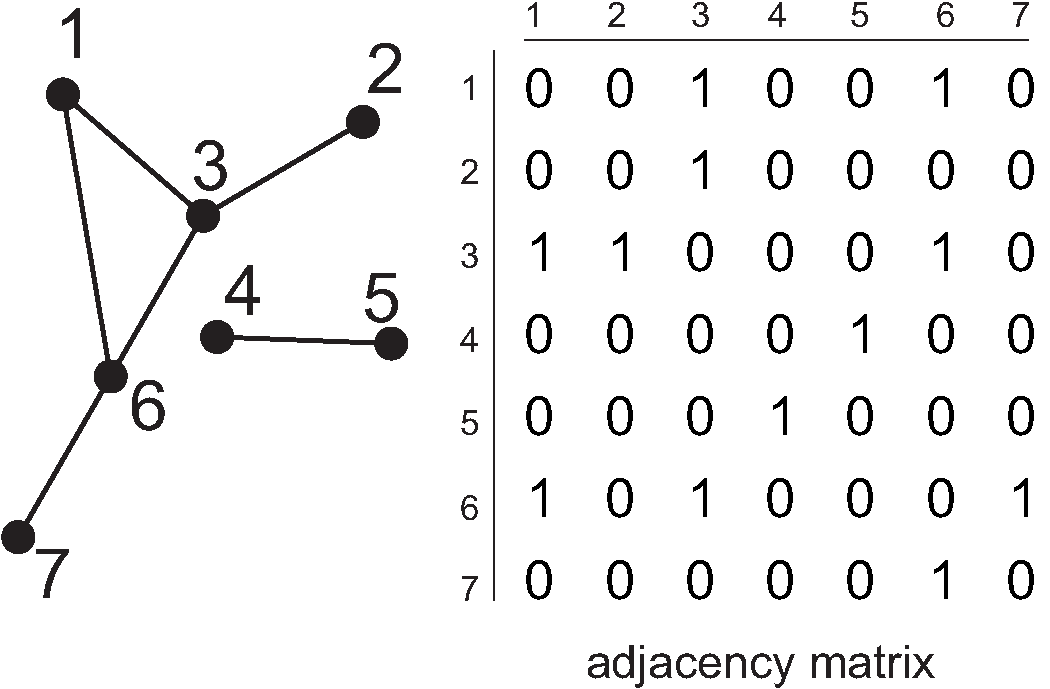} \hspace{0.3cm}
\includegraphics[width=5.2cm]{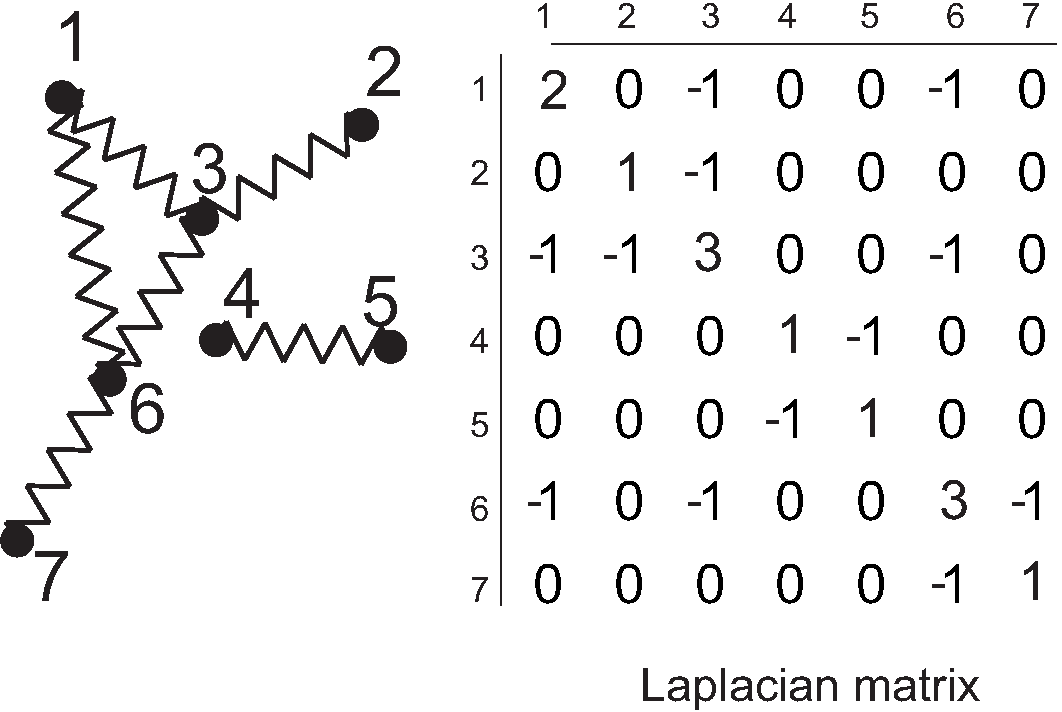}}
\caption{(a) Sample of a topological graph (collection of subgraphs) and its adjacency matrix;
(b) Elastic network corresponding to (a) and its Laplacian matrix.}
\label{fig21}
\end{figure}

Consider the polymer network as a graph or a collection of graphs, see \fig{fig21}a. Enumerate the
monomers of the $N$-atomic macromolecule by the index $i=1 \ldots N$. The adjacency matrix $A =
\{a_{ij}\}$ describing the topology (connectivity) of a polymer molecule is symmetric
($a_{ij}=a_{ji}$). Its matrix elements $a_{ij}$ take binary values, 0 and 1, such that the diagonal
elements vanish, i.e. $a_{ii}=0$, and for off-diagonal elements, $i\ne j$, one has $a_{ij}=1$, if
the monomers $i$ and $j$ are connected, and $a_{ij}=0$ otherwise.  The Laplacian matrix
$L=\{b_{ij}\}$ is by definition as follows: $b_{ij} = -a_{ij}$ for $i\ne j$, and $b_{ii} =
\sum_{j=1}^{N}a_{ij}$, as shown in \fig{fig21}b, i.e. $L=d I-A$, where $d$ is the vector of vertex degrees of the graph and $I$ is the identity matrix. The eigenvalues $\lambda_n$ ($n=1,...,N$) of the symmetric matrix $L$ are real. For regular graphs (i.e for graphs with constant vertex degrees) the spectra of the adjacency and Laplacian matrices are uniquely connected to each other.

In physical literature the spectrum of adjacency matrix is interpreted as the set of resonance
frequencies, while the Laplacian spectrum defines the relaxation of the system. Thus, measuring the
response of the diluted solution of noninteracting polymer molecules on external excitations with
continuously changing wavelength, we expect to see the signature of eigenmodes in the spectral
density as peaks at some specific frequencies. Some other applications of adjacency and Laplacian
matrices in graph theory and optimization are thoroughly described in \cite{mohar1997some} and
\cite{chung1997spectral}.

In the Ref. [\refcite{kovaleva}] we investigate the eigenvalue statistics of exponentially weighted ensembles of full binary trees (known as ``dendrimers'' in macromolecular physics) and $p$-branching star graphs. We show that spectral densities of corresponding adjacency matrices demonstrate peculiar ultrametric structure inherent to sparse systems and ensembles of linear chains exponentially weighted in their lengths. In particular, the tails of the distribution for binary trees again share the Lifshitz singularity discussed througout the text, while the spectral statistics of $p$-branching star-like graphs is less universal, being strongly dependent on $p$. 

The hierarchical structure of spectra of adjacency matrices is interpreted as sets of resonance frequencies, that emerge in ensembles of dendrimers. However, the this spectrum is not determined by the cluster topology, but rather has the number-theoretic origin, reflecting the peculiarities of the rare-event statistics typical for one-dimensional systems with a quenched structural disorder. The similarity of spectral densities of an individual dendrimer and of ensemble of linear chains with exponential distribution in lengths, demonstrates that dendrimers could be served as simple disorder-less toy models of one-dimensional systems with quenched disorder.

\subsection{Ultrametricity and chaotic Hamiltonian systems}

At the end of this review, I would like to mention the estimation of the survival probability in the chaotic Hamiltonian systems [\refcite{chaotic_avet}]. We considered the dynamical system described by the area-preserving standard mapping
\be
\left\{\begin{array}{l}
y_{t+1} = y_t - K/(2\pi) \sin 2\pi x_t \medskip \\
x_{t+1} = x_t + y_{t+1} \quad \mod 1
\label{eq:rec}
\end{array}\right.
\ee
It is known that for $K\in [0, K_g[$ the phase space of the system described by \eq{eq:rec} has disjoined islands of integrability which is destroyed as $K$ approaches $K_g$ from below, where $K_g\approx 0.97163540631$ (see Ref. [\refcite{mackay}]). Above the critical value $K_g$ the behavior of the system is less universal: many invariant Kolmogorov-Arnold-Moser (KAM) tori disappear, but still some islands of metastability survive around the biggest resonances.

It is known for this system that $P(t)$, the normalized number of recurrences staying in some given domain of the phase space at time $t$ (so-called ``survival probability'') has the power-law asymptotics, $P(t)\sim t^{-\nu}$. In the Ref. [\refcite{chaotic_avet}] we have presented semi-phenomenological arguments which enable us to map the dynamical system \eq{eq:rec} near the chaos border onto the effective ``ultrametric diffusion'' on the boundary of a tree-like space with hierarchically organized transition rates. Our consideration is based only on measurable ``macroscopic'' characteristics acquired in course of the iteration of the map \eq{eq:rec}. To be precise, we relied on the following well-established and confirmed facts: i) the number of principal resonances follows the Fibonacci sequence when $K\nearrow K_g$ [\refcite{hanson,chir2}]; ii) the generic behavior of phase trajectories is as follows: the phase trajectory stays in the vicinity of some resonance (low-flux ``cantori'') and then rapidly crosses the chaotic sea until another metastable low-flux cantori is reached [\refcite{hanson,mackay2}]; iii) the phase space of the standard mapping is self-similar being usually represented by a binary (i.e. 3-branching) Cayley tree [\refcite{hanson,meiss1,meiss2}]; iv) the survival probability has power-law asymptotic behavior, $P(t)\sim t^{-\nu}$ (for the first time this has been shown in Ref. [\refcite{karney1,chir1}]).

In Ref. [\refcite{chaotic_avet}] we have estimated the exponent $\nu$ as
\be
\nu=\ln 2/\ln (1+r_g)\approx 1.44
\ee
where $r_g=(\sqrt{5}-1)/2$ is the critical rotation number. The survival probability, $P(t)$, is the normalized number of recurrences \eq{eq:rec} which stay in some given domain of the phase space at time $t$. Different research groups present different arguments for estimates of $\nu$, typically $1<\nu \lesssim 3$. The most intriguing contradiction concerns the discrepancy in reported values of $\nu$. The numerical simulations [\refcite{chir1,chir3}] demonstrate $\nu\approx 1.4 \div 1.5$, while almost all known analytic constructions give essentially larger exponents: $\nu\approx 1.96$ in Ref. [\refcite{meiss1}] and $\nu\approx 3.05$ in Ref. [\refcite{meiss2}]. The scaling analysis [\refcite{chir2}] valid just near the chaos boundary gives $\nu=3$. The special attention should be paid to the recent works [\refcite{ketzm,ven1,ven2}]. In Ref. [\refcite{ketzm}] the authors demonstrate that by an appropriate randomization of the ``Markov tree model'' proposed in Ref. [\refcite{meiss1,meiss2}] one can arrive at the value $\nu\approx 1.57$. The works [\refcite{ven1,ven2}] claim $\nu=3$ for sticking of trajectories near $K_g$ in agreement with [\refcite{chir2}] and $\nu=3/2$ for trapping of chaotic trajectories in the vicinity of cantori for $K\approx 2\ell/\pi$ (where $\ell$ is a nonzero integer). The similar exponent, $\nu=3/2$, was also obtained for a standard map in the work [\refcite{verg}]. There is a point of view that the value $\nu\approx 1.4 \div 1.5$ corresponds to an intermediate behavior of the system which is not yet reached the stationary regime -- see the corresponding discussion in Ref. [\refcite{chir2,weiss, chir4,art}]. In Ref. [\refcite{chaotic_avet}] we have presented some simple arguments in favor of the statement that the value $\nu\approx 1.4 \div 1.5$ could be the actual decay exponent of the survival probability $P(t)$ for $t\to\infty$ related to the diffusion in the ultrametric landscape.

I would like to conclude by saying that the conjectured approach offers a possibility to rise some
interesting questions concerning the internal structure of the standard mapping \eq{eq:rec}. For
example, it would be desirable to check numerically the existence of the ``secondary'' Fibonacci
sequences in addition to the standard ones (see Ref. [\refcite{chaotic_avet}] for details). In case of their existence, one would be intriguing to think about the phyllotaxis in chaotic dynamical systems. By this conjecture we would like to attract the attention of researchers working in nonlinear dynamical and chaotic systems to the language developed for the description of stochastic processes in ultrametric spaces.

\begin{acknowledgments}
I want to express my sincere gratitude to my collaborators and colleagues with whom I have worked and/or discussed over years the topics reflected in this review: Vladik Avetisov, Alexander Gorsky, Alexander Grosberg, Yan Fyodorov, Pavel Krapivsky, Leonid Pastur, Kirill Polovnikov, Michael Tamm, Olga Valba, and Raphael Voituriez. I highly appreciate fruitful comments received in different years from Arezki Bodaoud, John Gemmer, Yuri Manin, Michael Marder, Leonid Mirny, Lev Truskinovsky, and Dima Shepelyansky. The supports of RFBR grant 16-02-00252 and of EU-Horizon 2020 IRSES project DIONICOS (612707) are highly acknowledged. I would cordially thank Yurij Holovatch for his kind proposal to write this Chapter.
\end{acknowledgments}

\end{document}